\documentclass[nofootinbib,superscriptaddress,10pt]{revtex4}

\usepackage{epstopdf}
 
\usepackage{amssymb}
\usepackage{epsfig}
\usepackage{subfigure}
\usepackage{graphicx}
\usepackage{longtable}
\usepackage{float}
\usepackage{pslatex}
\usepackage{hyperref}

\begin{document}

\newcommand{\dd}{\mathrm{d}}
\newcommand{\ii}{\mathrm{i}}
\newcommand{\dis}{\displaystyle}
\newcommand{\bupup}{\vspace*{-0.5\baselineskip}}
\newcommand{\bup}{\vspace*{-\baselineskip}}

\newcommand{\sqrtsnn}{\sqrt{s_{_{NN}}}}
\def\mean#1{\ensuremath{\left<#1\right>}}
\newcommand{\pom}{I\!\!P}
\newcommand{\gaga}{\gamma\,\gamma}
\newcommand{\gp}{\gamma\,p}
\newcommand{\gA}{\gamma\,A}

\def\AuAu{Au+Au}
\def\PbPb{Pb+Pb}
\def\pp{p+p}

\def\pT{\mbox{$p_{T}$}}
\def\ee{\mbox{$e^+e^-$}}

\def\sqrtsNN{\mbox{$\sqrt{s_{NN}}$}}

\def\eq#1{{Eq.~(\ref{#1})}}
\def\tab#1{{Table~\ref{#1}}}
\def\fig#1{{Figure~\ref{#1}}}
\def\order#1{\mathcal{O}{(#1)}}

\providecommand{\jpsi}{J/\psi}
\providecommand{\psip}{\psi^{\prime}}
\providecommand{\ups}{\Upsilon}
\providecommand{\upsp}{\Upsilon^{\prime}}
\providecommand{\upspp}{\Upsilon^{\prime\prime}}
\providecommand{\dzero}{$D^0$}
\providecommand{\dplus}{$D^+$}
\providecommand{\dstar}{$D^{*+}$}

\providecommand{\mean}[1]{\ensuremath{\left<#1\right>}}

\newcommand{\ncoll}{N_{\rm coll}}
\newcommand{\npart}{N_{\rm part}}
\newcommand{\raa}{R_{\rm AA}}
\newcommand{\pt}{p_{\rm T}}

\newcommand{\ch}{\chi^{2}}
\newcommand{\chndf}{\chi^{2}/ndf}
\newcommand{\dch}{\Delta \chi^{2}}

\newcommand{\mee}{m_{e^+ e^-}}
\newcommand{\mlee}{m_{e^{\pm} e^{\pm}}}

\DeclareGraphicsExtensions{.eps,.jpg,.pdf,.png,.gif}

\title{Reference heavy flavour cross sections in pp collisions at $\sqrt{s}=2.76~\rm TeV$, using  a pQCD-driven $\sqrt{s}$-scaling of ALICE measurements at $\sqrt{s}=7~\rm TeV$}

\author{R.~Averbeck}
\affiliation{Research Division and ExtreMe Matter Institute EMMI, GSI Helmholtzzentrum f\"ur Schwerionenforschung, Darmstadt, Germany}
\author{N. Bastid}
\affiliation{LPC, Clermont-Ferrand, France}
\author{Z. Conesa del Valle}
\affiliation{CERN, Geneva, Switzerland}
\author{P. Crochet}
\affiliation{LPC, Clermont-Ferrand, France}
\author{A. Dainese}
\affiliation{INFN -- Sezione di Padova, Padova, Italy}
\author{X. Zhang}
\affiliation{LPC, Clermont-Ferrand, France}
\affiliation{CCNU, Wuhan, China}


\begin{abstract}
\noindent  
We provide a reference in proton--proton collisions at the energy of the Pb--Pb 2010 run
at the LHC, $\sqrt{s}=2.76$~TeV, for the $p_{\rm t}$-differential production cross section of 
$D^0$, $D^+$, and $D^{*+}$ mesons in $|y|<0.5$, of electrons from heavy flavour decays
in $|y|<0.9$, and of muons from heavy flavour decays in $2.5<y<4$. The reference is obtained
by applying a pQCD-driven scaling (based on the FONLL calculation) 
to ALICE preliminary data at $\sqrt{s}=7$~TeV. In order to validate the procedure, we 
scale the D meson cross section to $\sqrt{s}=1.96$~TeV and compare to the corresponding
measurements from the CDF experiment.
\end{abstract}


\maketitle

\thispagestyle{empty}

\clearpage

\setcounter{page}{1}
\section{Introduction}

In 2010, the Large Hadron Collider (LHC) delivered large samples of proton--proton (pp)
collisions at $\sqrt{s}=7$~TeV and of Pb--Pb collisions at $\sqrt{s_{NN}}=2.76$~TeV.
A pp reference at $\sqrt{s}=2.76$~TeV is required,
in order to compare heavy flavour production in Pb--Pb and pp collisions
via the nuclear modification factor of the $p_{\rm t}$ distributions
\begin{equation}
R_{\rm AA}(p_{\rm t})=\frac{1}{\langle T_{\rm AA}\rangle}\frac{{\rm d} N_{\rm AA}/{\rm d} p_{\rm t}}{{\rm d} \sigma_{\rm pp}/{\rm d} p_{\rm t}}\,.
\end{equation}
Here, 
$\langle T_{\rm AA}\rangle$ is the average value of the nuclear overlap function
in a given Pb--Pb centrality class, $N_{\rm AA}$ is the production 
yield per event of the considered
particle in that class, and $\sigma_{\rm pp}$ is its production cross section in pp collisions
at the same energy. 

In this note, we show that state-of-the-art perturbative QCD calculations provide 
an accurate guidance for extrapolating to lower energy 
the $p_{\rm t}$-differential cross sections measured at 7~TeV. It was already shown~\cite{PPRvol2}, using the MNR~\cite{MNR} NLO pQCD calculations at the energies $\sqrt{s}=5.5$ and $14$~TeV, that, despite the large 
spread for the cross section at a given energy with different values of the heavy quark masses
and factorization/renormalization scales, the ratio of the cross sections at the two energies
is much less dependent on the choice of the calculation parameters. 
We now use
the Fixed Order Next-to-Leading Log (FONLL) calculations~\cite{FONLL}, 
and apply the resulting scaling factor to preliminary ALICE 
cross section measurements  at 7~TeV: $D$ mesons ($D^0$, $D^+$, and $D^{*+}$) at mid-rapidity 
and single leptons (electrons at mid-rapidity and muons at forward rapidity) from 
charm and beauty hadron decays. 
 In order to validate the procedure, we 
scale the $D$ meson cross sections to $\sqrt{s}=1.96$~TeV and compare to the corresponding
measurements from the CDF experiment~\cite{CDFdata}.

After reporting the ALICE heavy flavour cross section measurements at $\sqrt{s}=7$~TeV (section~\ref{sec:data_analysis}), we describe (section~\ref{sec:recipe}) the procedure adopted
for the energy scaling, and (section~\ref{sec:results}) the scaling factors and 
resulting cross sections at 2.76~TeV. 

\section{ALICE heavy-flavour production measurements at $\sqrt{s}=7~{\rm TeV}$ }
\label{sec:data_analysis}

Here we briefly present the ALICE measurements at $\sqrt{s}=7~$TeV that will be used for the scaling.

\begin{figure}[!htbp]
\begin{center}
        \includegraphics[width=0.45\columnwidth]{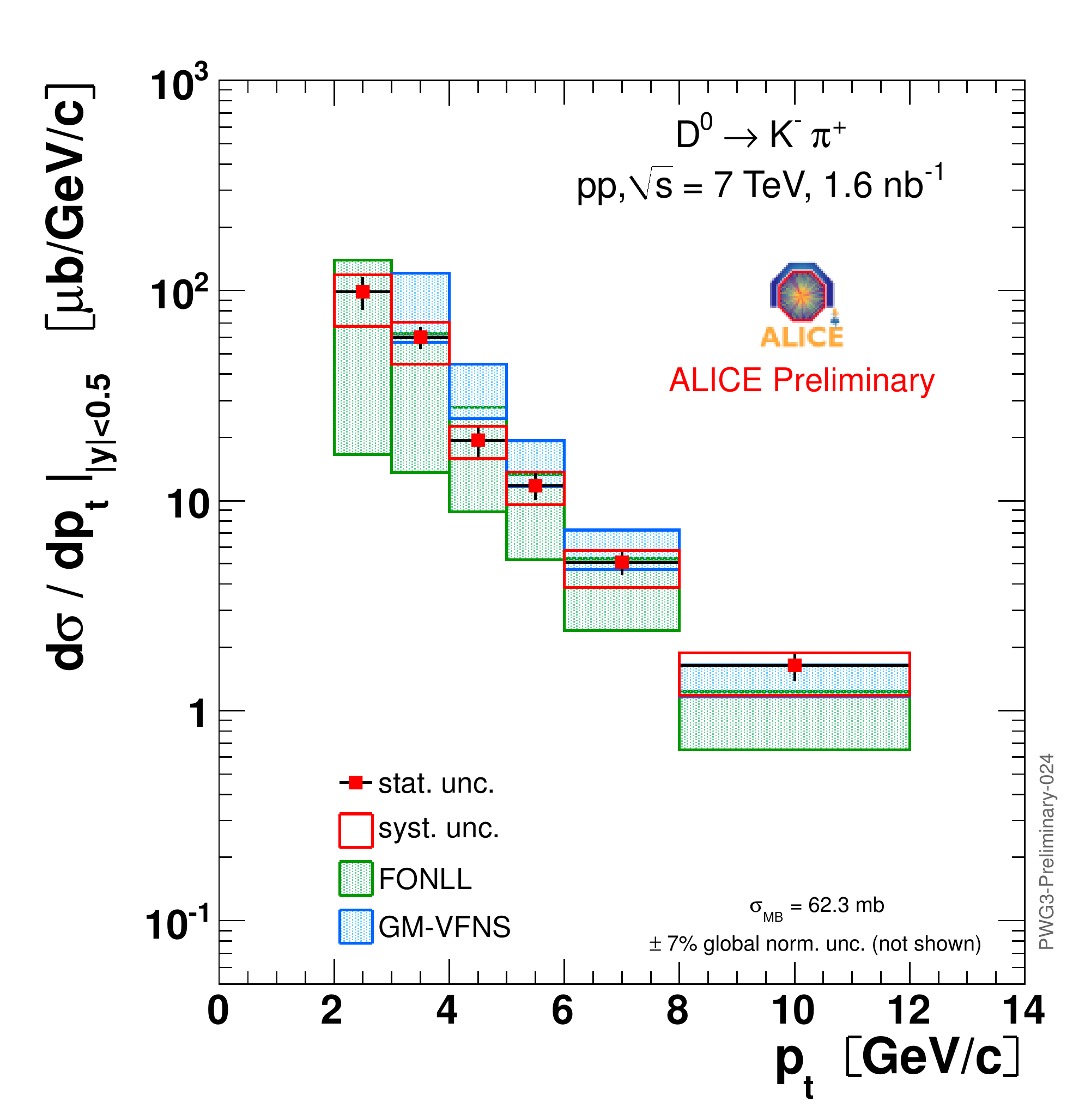}
        \includegraphics[width=0.45\columnwidth]{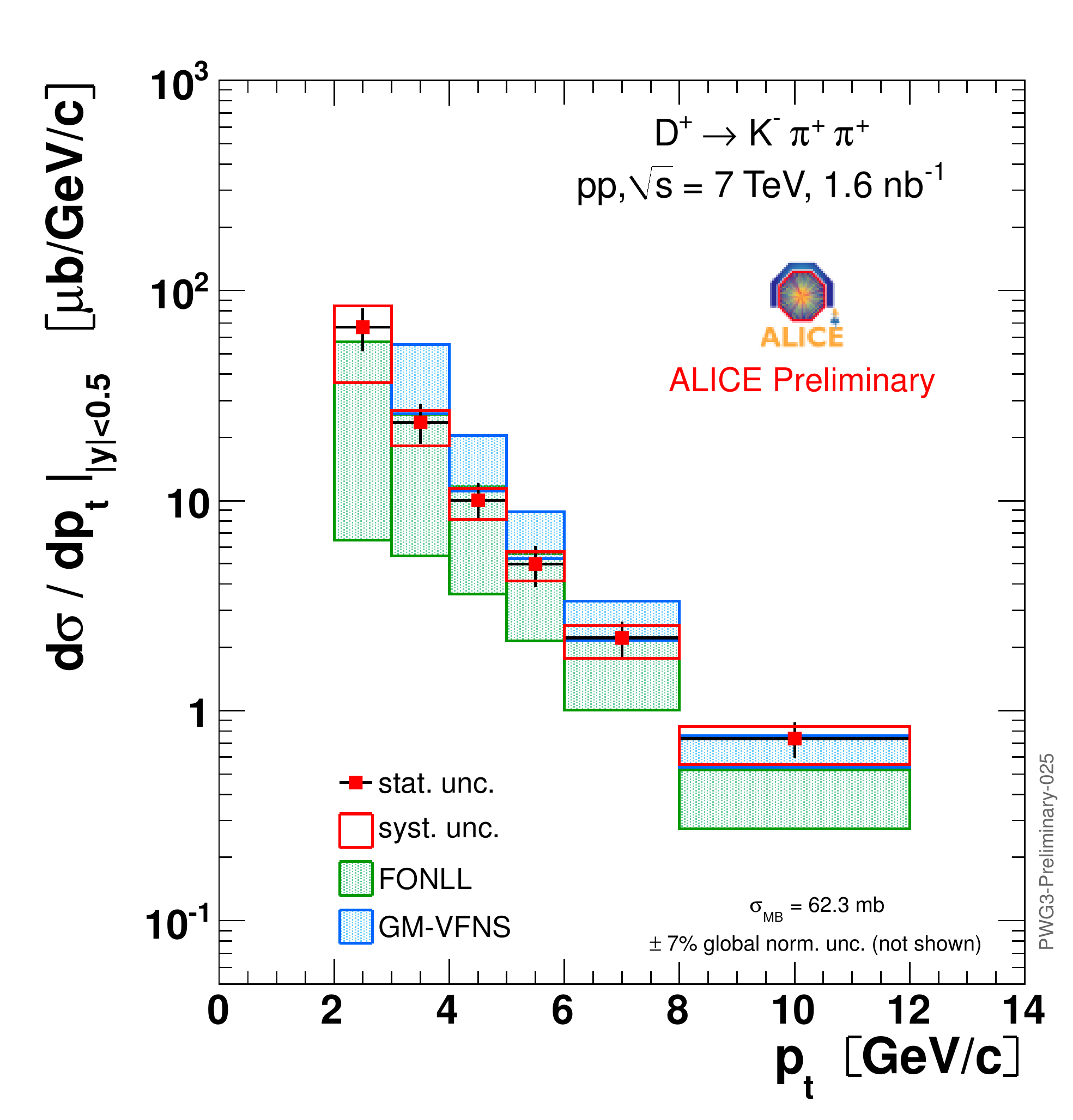}
        \includegraphics[width=0.45\columnwidth]{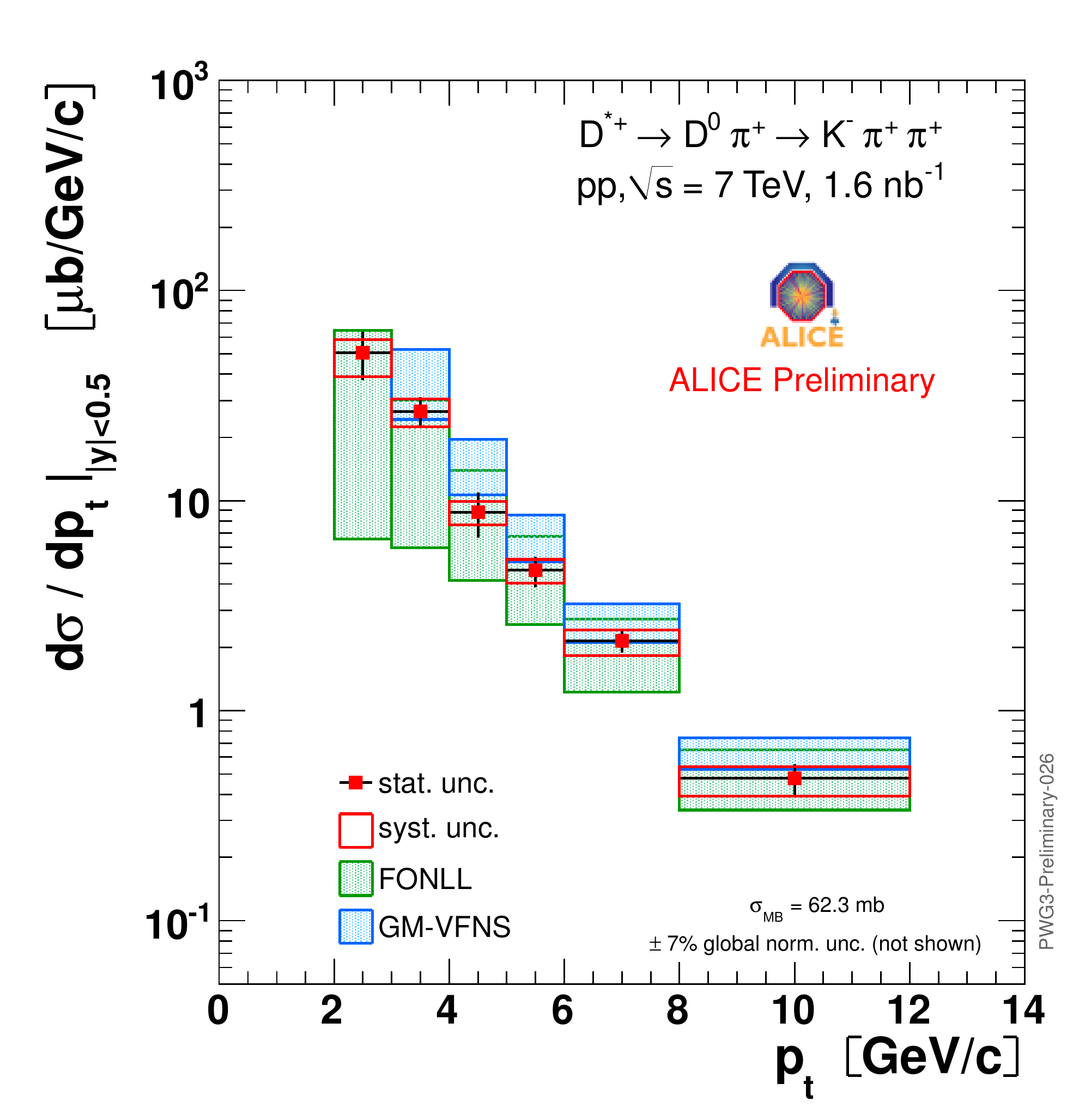}
\caption{ALICE $D^0$, $D^+$, and \dstar~$p_t$ differential preliminary cross sections in $|y|<0.5$
at $\sqrt{s}=7$~TeV~\cite{andrea}. The FONLL~\cite{FONLL,Cacciari} and GM-VFNS~\cite{GMVFNS} predictions are compared to the data.}
\label{fig:DPreliminary}
\end{center}
\end{figure}

The preliminary results on the \dzero, \dplus, and \dstar~cross sections at $\sqrt{s}=7~$TeV, 
measured at central rapidity ($|y|<0.5$) using the decay channels
$D^0 \rightarrow K^{-} \, \pi^{+}$,
$D^+ \rightarrow K^{-} \, \pi^{+} \, \pi^{+}$,
$D^{*+} \rightarrow  D^0 \, \pi^{+}  \rightarrow K^{-} \, \pi^{+} \, \pi^{+} $, 
are shown in Fig.~\ref{fig:DPreliminary}.
The data are compared to pQCD predictions based on the 
FONLL~\cite{FONLL,Cacciari} and GM-VFNS~\cite{GMVFNS} calculations.
These results are described in~\cite{andrea,renu}. 
Let us recall that these measurements correspond to the direct charm production, as they were corrected for the B-mesons feed-down contribution. 

\begin{figure}[!htbp]
\begin{center}
        \includegraphics[width=0.475\columnwidth]{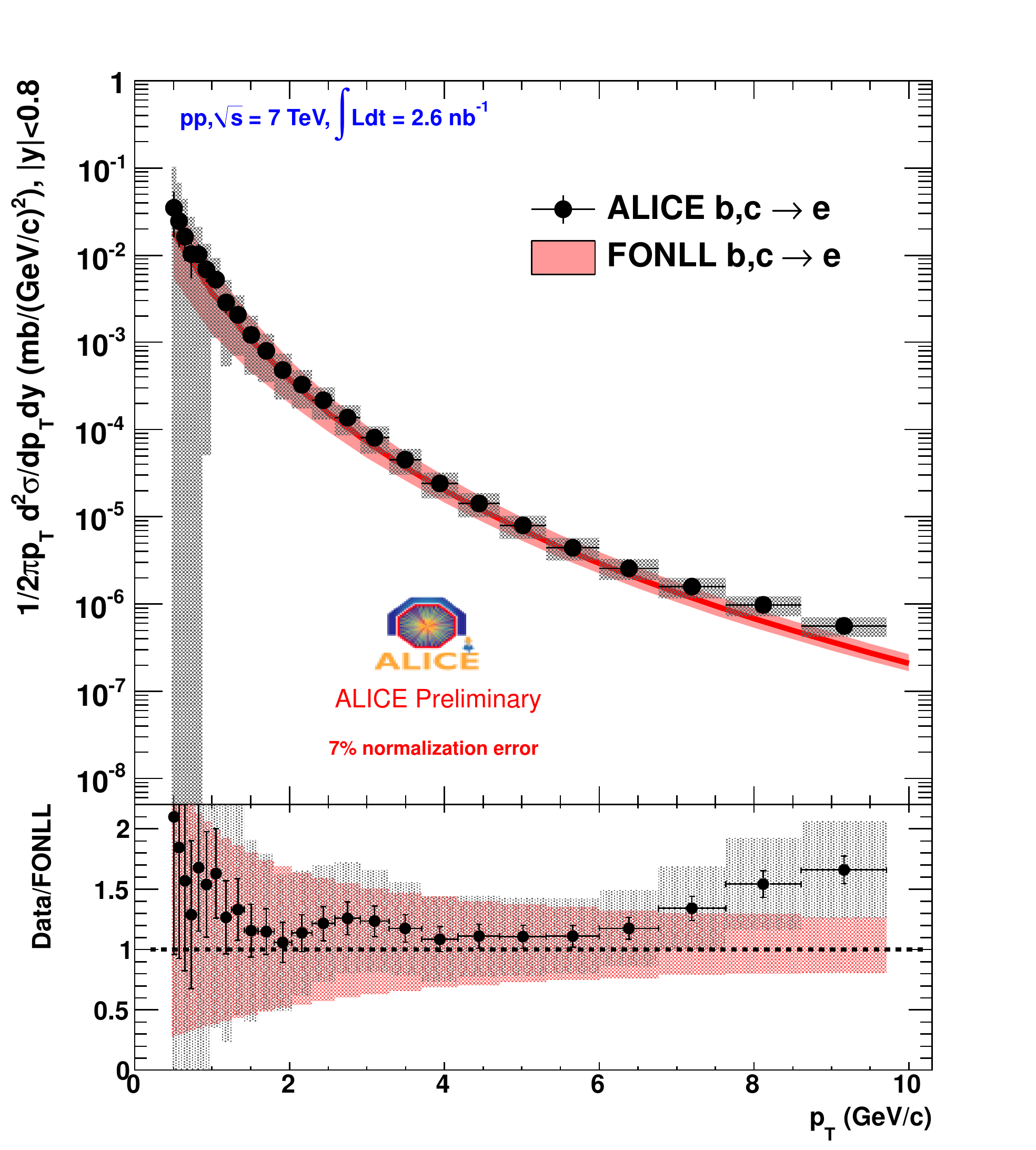}
\caption{ALICE heavy flavour decay electron $p_t$ differential production cross section for pp collisions at $\sqrt{s} = 7$~TeV~\cite{noteHFE}. 
The FONLL pQCD calculation~\cite{FONLL,Cacciari} is compared to the data.}
\label{fig:SingleElectronsPreliminary}
\end{center}
\end{figure}

Fig.~\ref{fig:SingleElectronsPreliminary} shows the preliminary cross section of 
electrons from heavy flavour decays for pp collisions at $\sqrt{s} = 7$~TeV
and the corresponding prediction from the FONLL pQCD 
calculation~\cite{FONLL,Cacciari}. A reasonable agreement between data and model
calculation is observed.
The analysis procedure is discussed in~\cite{noteHFE}.

\begin{figure}[!htbp]
\begin{center}
        \includegraphics[width=0.9\columnwidth]{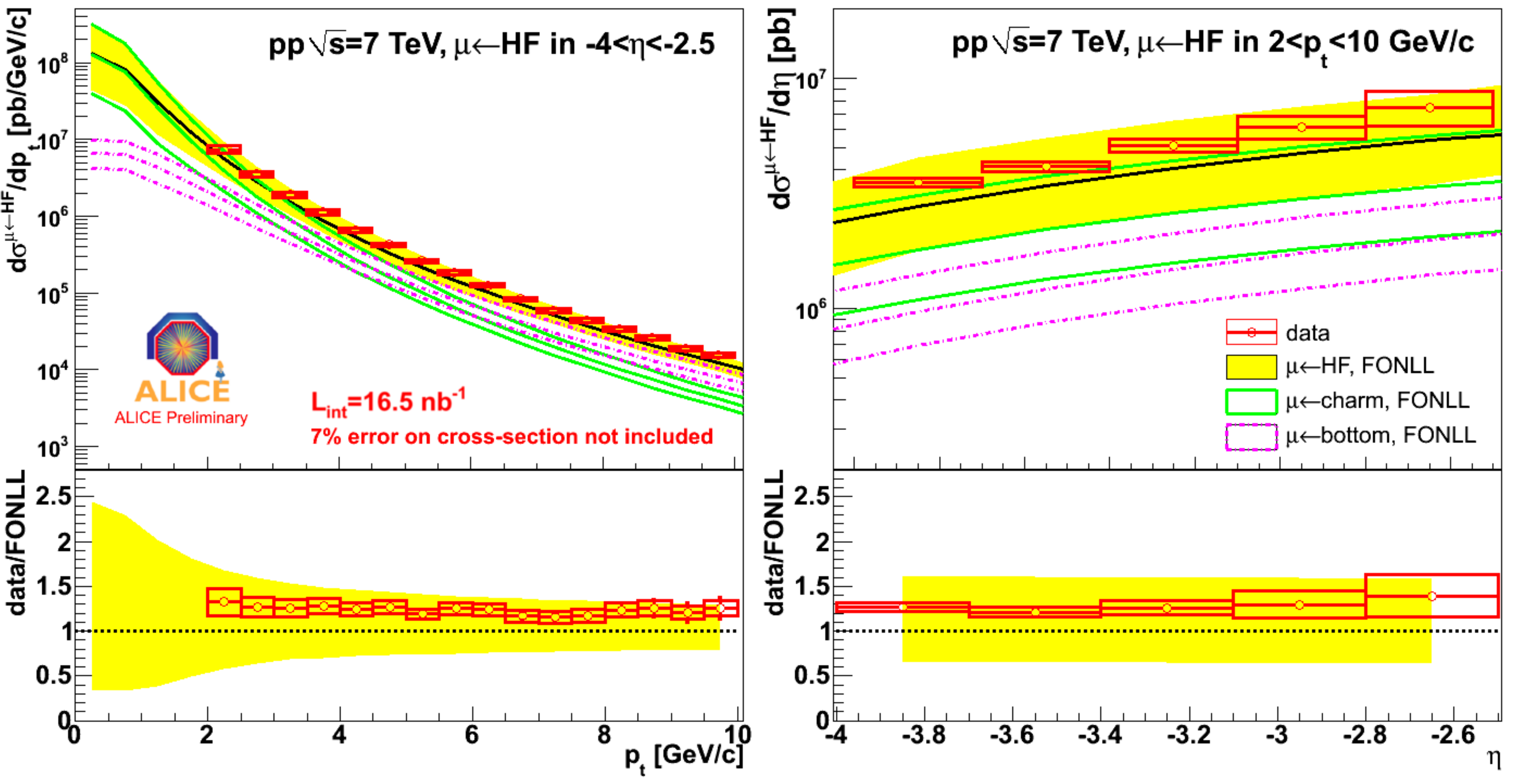}
\caption{ALICE $p_{\rm t}$ and $\eta$ differential production cross section of muons from 
heavy flavour decays, in $2.5<y<4$, in pp collisions at $\sqrt s$ = 7 TeV (symbols)~\cite{XZnote}. 
The results are compared to FONLL predictions~\cite{FONLL,Cacciari}.}
\label{fig:SingleMuonPreliminary}
\end{center}
\end{figure}

The inclusive $p_{\rm t}$ and $\eta$ differential cross section of muons from heavy flavour 
decays, in the rapidity range $2.5<y<4$, in pp collisions 
at $\sqrt s$~=~7 TeV is displayed in Fig.~\ref{fig:SingleMuonPreliminary}
~\cite{XZnote}, along with the corresponding FONLL prediction~\cite{FONLL,Cacciari}.

\section{Energy scaling procedure} 
\label{sec:recipe}

In order to scale the ALICE 7~TeV cross-sections to a given energy we consider the scaling factors provided by different theoretical calculations. 
The {\sf FONLL}~\cite{FONLL,Cacciari} driven scaling is set as our reference scaling and is evaluated considering the different sets of scales (factorization scale $\mu_F$, renormalization scale $\mu_R$) and quark masses ($m_c$ and $m_b$). We consider the standard parameter variations
that are used to evaluate the theoretical uncertainty on the charm and beauty production cross 
sections (see e.g.~\cite{baines,vogt}):
\begin{itemize}
\item $0.5<\mu_F/\mu_0<2$ (central value: 1);
\item $0.5<\mu_R/\mu_0<2$ (central value: 1);
\item with the constraint $0.5<\mu_F/\mu_R<2$;
\item $1.3<m_c<1.7~$GeV (central value: 1.5~GeV) and $4.5<m_b<5.0~$GeV (central value: 4.75~GeV);
\end{itemize}
where $\mu_0=\sqrt{m_{\rm Q}^2+p_{\rm t,Q}^2}= m_{\rm t,Q}$.

The procedure to compute the {\sf FONLL} scaling factor from 7~TeV to an energy of $\alpha$~TeV is: 
 \begin{enumerate}
 \item Rebin the FONLL predictions for $\sigma (\alpha) $ and $ \sigma (7) $ for the different sets of scales ($\mu_F$, $\mu_R$), and quark masses ($m_c$ and $m_b$) according to the ALICE 7~TeV $p_{\rm t}$ binning for each observable. 
 \item Estimate the FONLL $\sigma (\alpha) / \sigma (7) $ ratio per observable\footnote{
 	This means that for single leptons this is the ratio of charm + bottom contributions. 
	}  considering that:
 \begin{itemize}
 \item The central value is the ratio of the central predictions at both energies and 
 \item its uncertainty is defined by the envelope (spread) of the ratio of the calculations for the different sets of parameters. 
 	Note that for a given quark flavour we can consider that the theoretical calculation parameters are correlated (equal) at different energies. However, we do not assume they are equal for charm and beauty.  
 \end{itemize}
 \item Multiply the ALICE 7~TeV cross-sections by the FONLL $\sigma (\alpha) / \sigma (7)$ binned ratio. 
 \item Propagate the uncertainties:
 \begin{itemize}
 \item on the FONLL ratios, 
 \item on the uncertainties of the 7~TeV measurement, 
 \item combine these uncertainties. 
 \end{itemize}
 \end{enumerate} 

\noindent
The considered cross-checks of the scaling procedure are :  
 \begin{enumerate}
 \item Interpolate to Tevatron energy ($\rm p\overline p$ at $\sqrt s=1.96$~TeV) to compare to the D meson measurements by the CDF Collaboration~\cite{CDFdata};
 \item Compare the scaling factor from {\sf FONLL} to that obtained from other (Fixed Order) 
 pQCD calculations (NLO {\sf MNR}~\cite{MNR}, {\sf GM-VFNS}~\cite{GMVFNS}). 
  \end{enumerate}

\section{Results} 
\label{sec:results}

%
\subsection{$D$ mesons} 
\label{sec:CharmScaling}

%
\subsubsection{Scaling factor to $\sqrt s=2.76$~TeV}
\label{sec:CharmScaling276TeV}

\begin{figure}[!htbp]
\begin{center}
        \includegraphics[width=0.47\columnwidth]{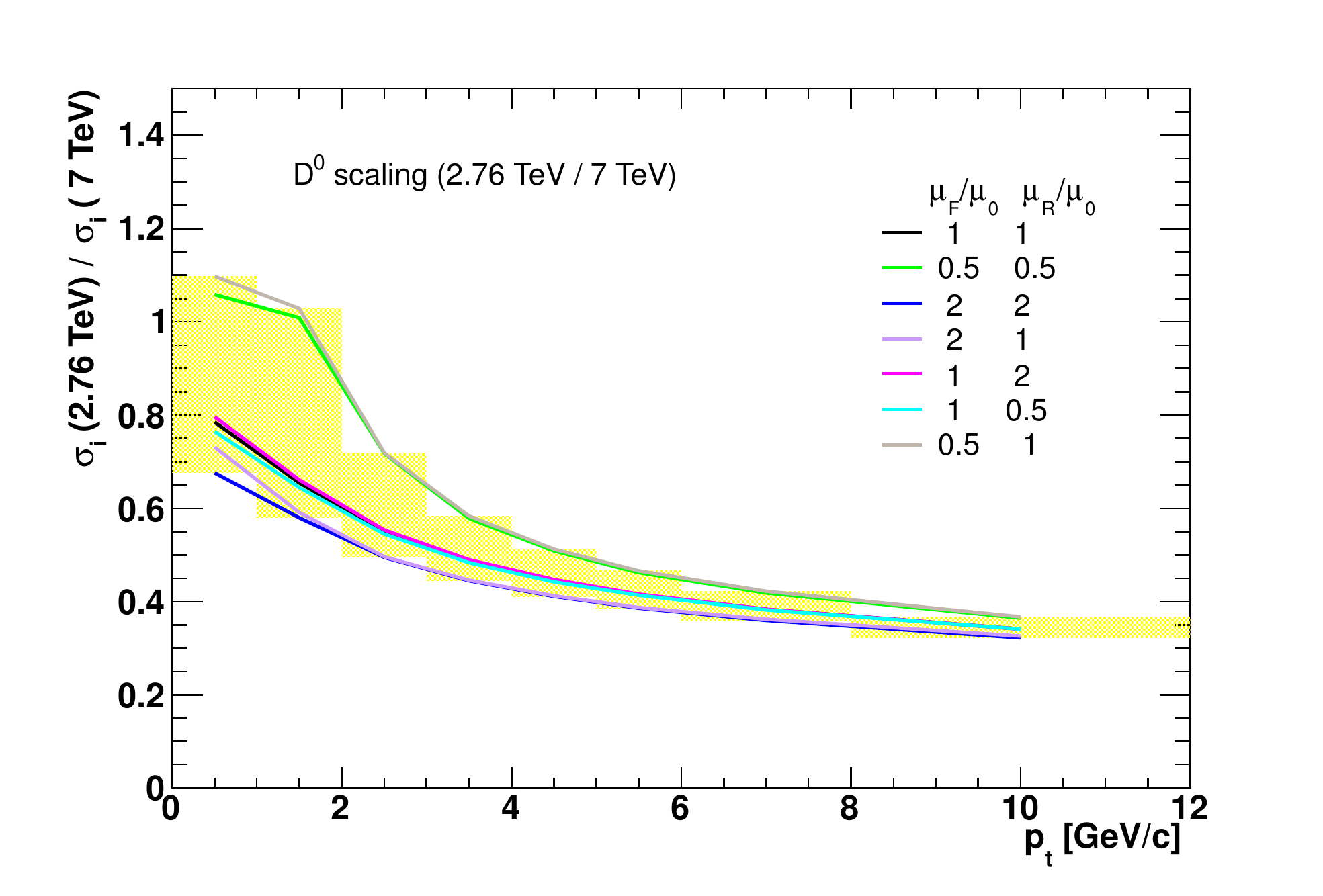}
        \includegraphics[width=0.47\columnwidth]{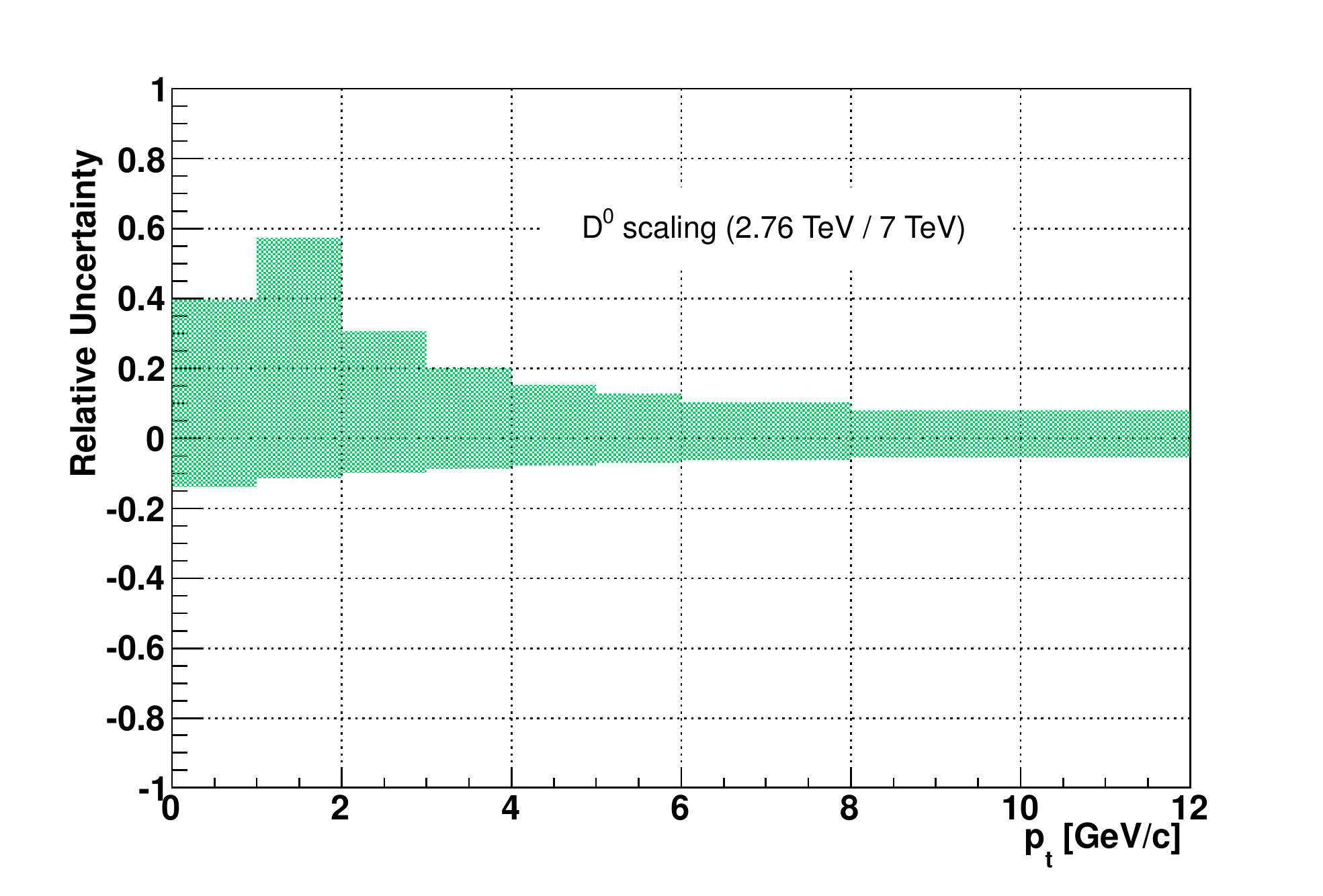}
\caption{\dzero~{\sf FONLL} scaling to 2.76~TeV from 7 TeV (7~TeV data binning).}
\label{fig:D0Scaling}
\end{center}
\begin{center}
        \includegraphics[width=0.47\columnwidth]{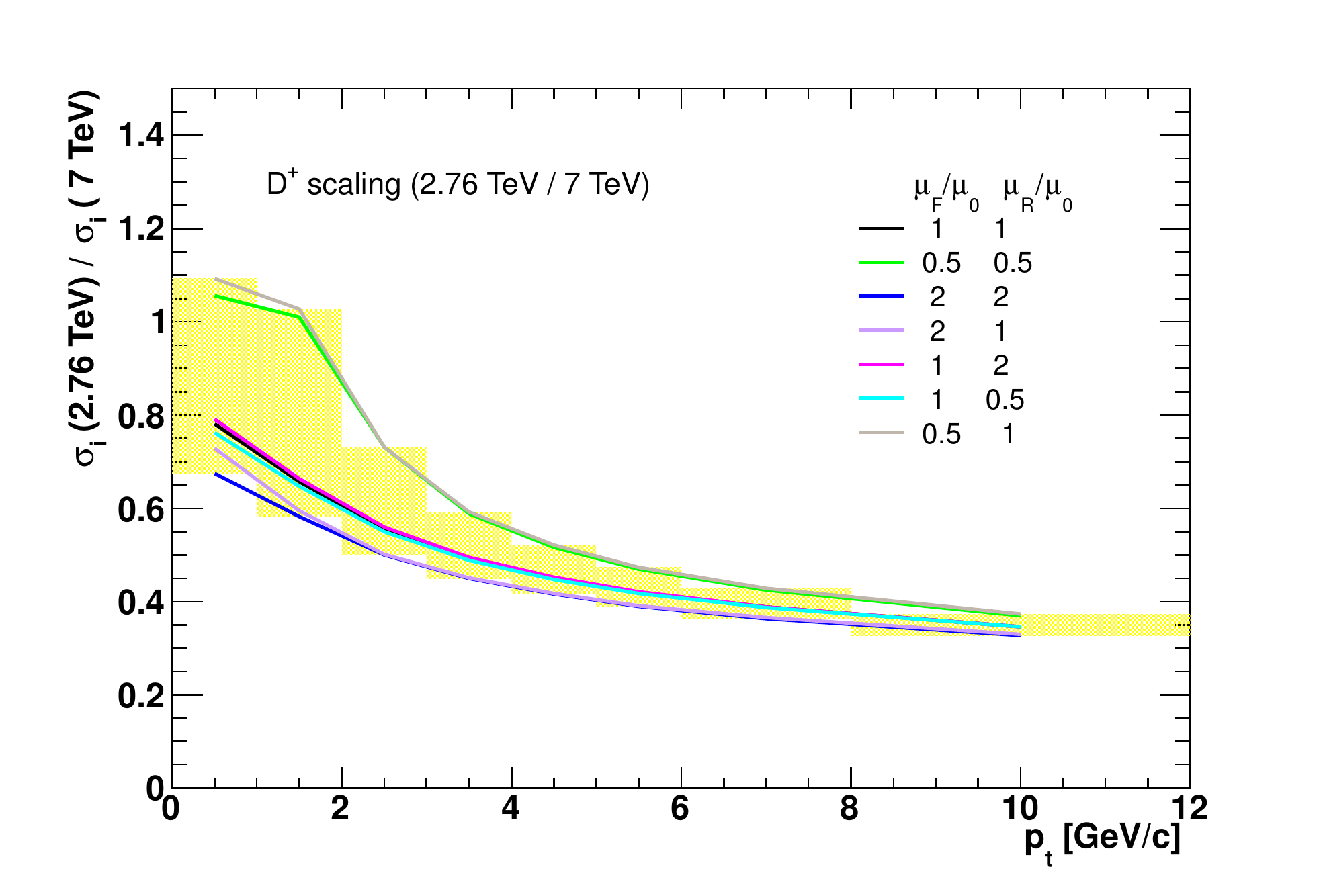}
        \includegraphics[width=0.47\columnwidth]{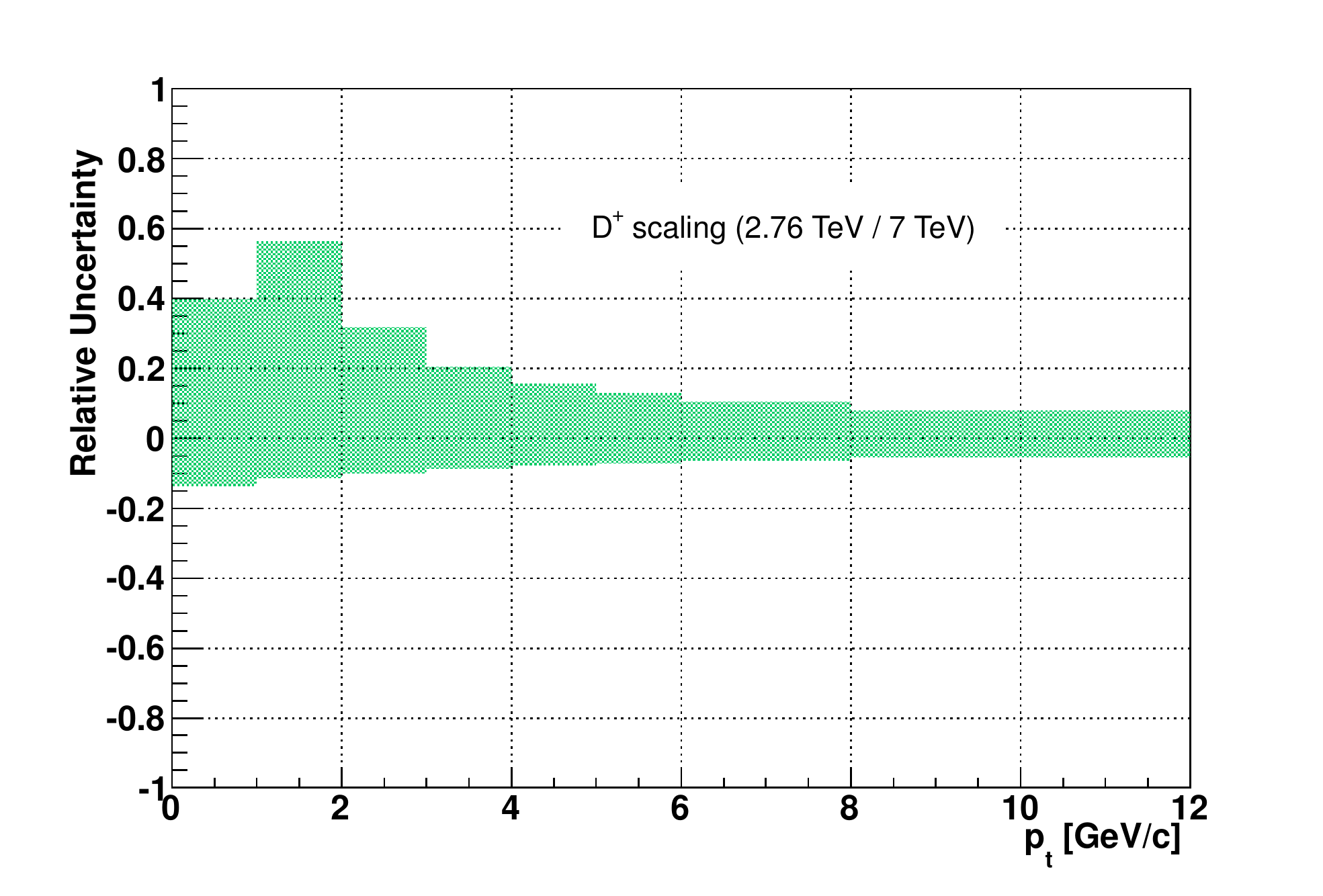}
\caption{\dplus~{\sf FONLL} scaling to 2.76~TeV from 7 TeV (7~TeV data binning).}
\label{fig:DplusScaling}
\end{center}
\begin{center}
        \includegraphics[width=0.47\columnwidth]{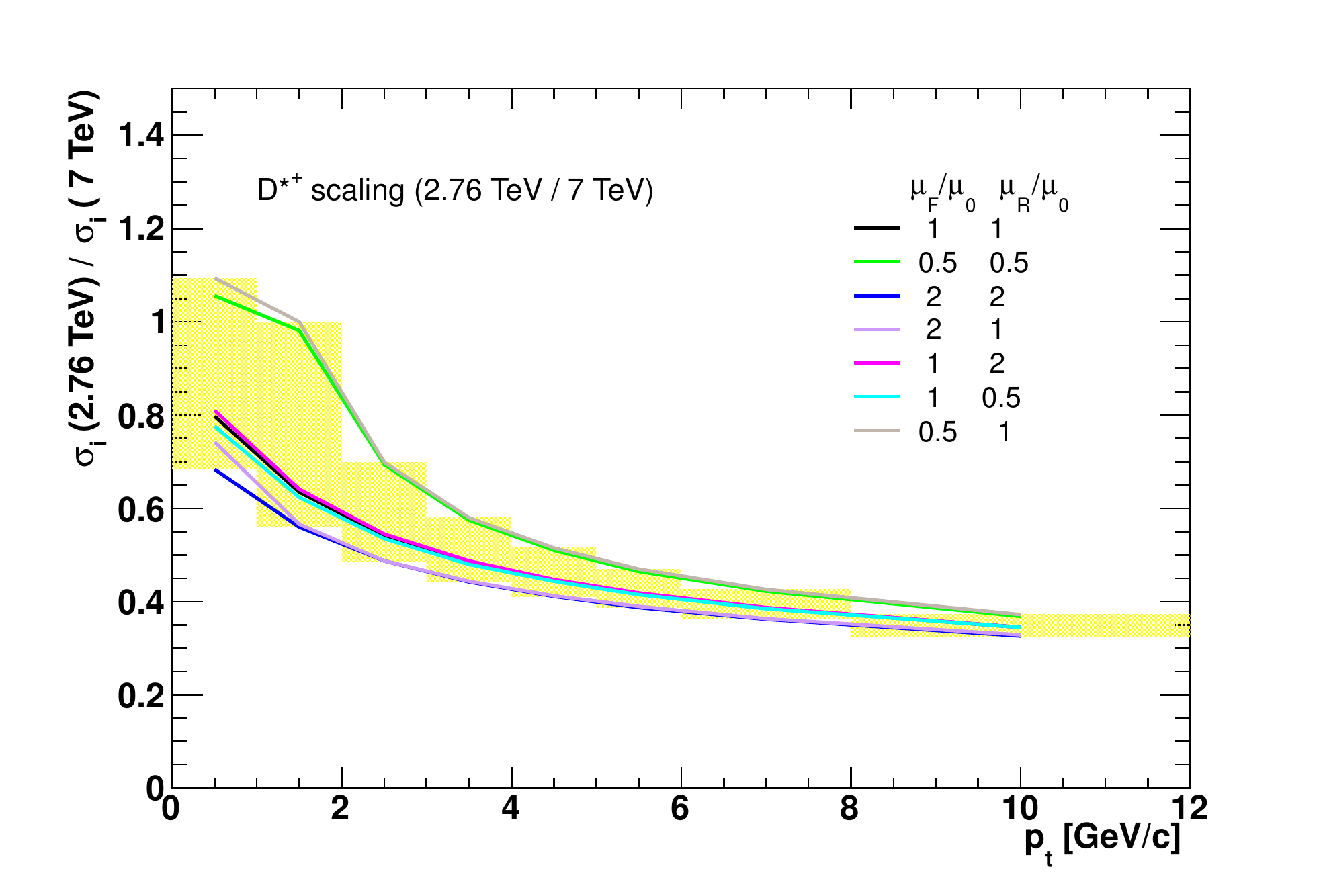}
        \includegraphics[width=0.47\columnwidth]{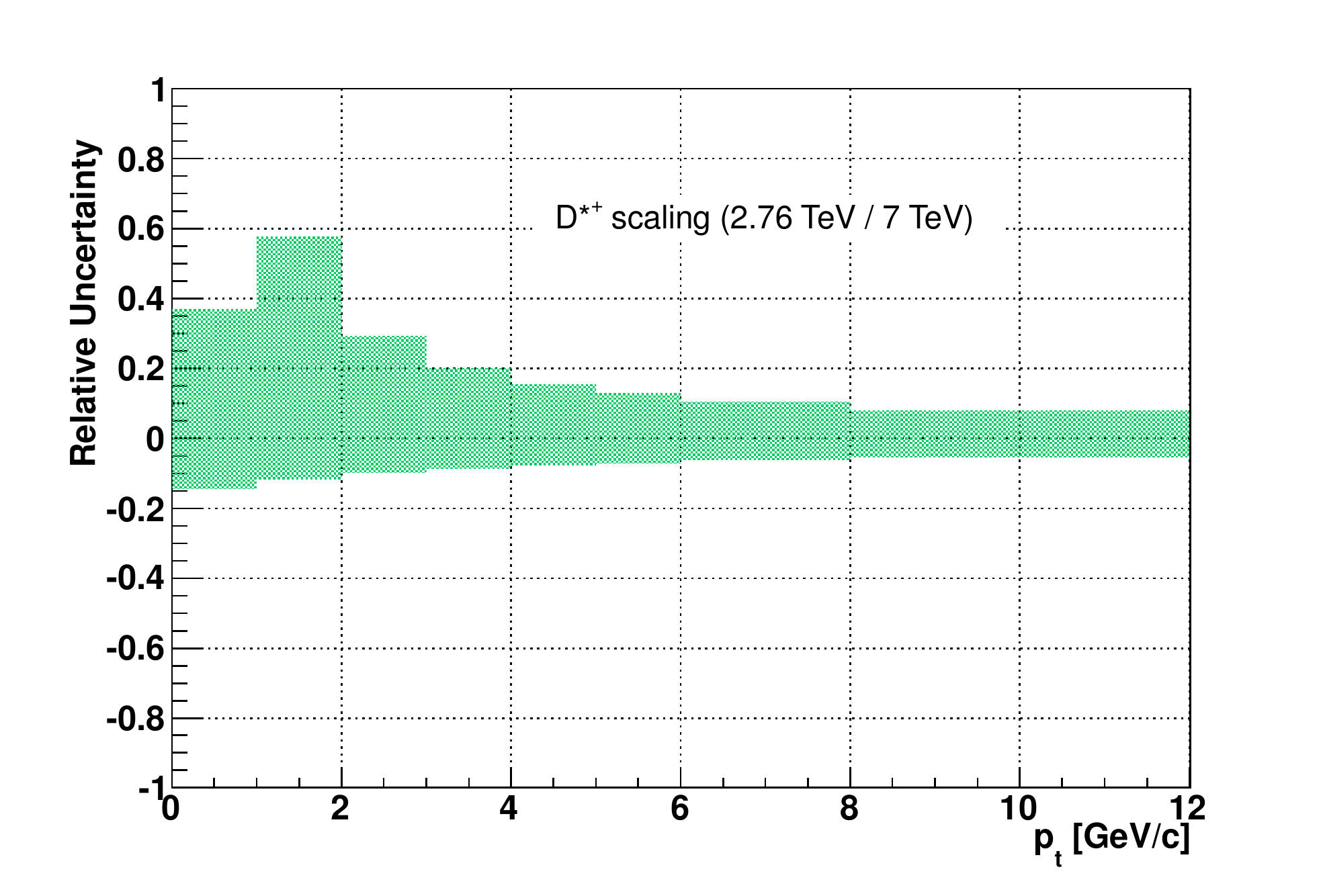}
\caption{\dstar~{\sf FONLL} scaling to 2.76~TeV from 7 TeV (7~TeV data binning).}
\label{fig:DstarScaling}
\end{center}
\end{figure}

The {\sf FONLL} scaling factors for \dzero, \dplus, and \dstar were calculated as described in the previous section and are shown in Figs.~\ref{fig:D0Scaling},~\ref{fig:DplusScaling}, and~\ref{fig:DstarScaling}~(left-hand panels) together with their respective relative uncertainties~(right-hand panels). The scaling factor obtained with the different sets of scales are drawn with solid lines, while the resulting global scaling is depicted by a yellow filled band. The central value of the scaling is obtained with $\mu_F/\mu_0=\mu_R/\mu_0=1$. The values of the scales for the other sets are
reported in the legend ($\mu_F/\mu_0$, $\mu_R/\mu_0$). 
We can observe that the scaling factor depends mainly on the value of the factorization scale, with almost no dependence on the renormalization scale.
This is due to the fact that, for the same heavy quark $p_{\rm t}$, different Bjorken $x$ 
ranges are probed at 2.76 and at 7 TeV, and changing the factorization scale affects the $x$ dependence of the parton distribution functions (PDFs).
The scaling factor does not depend on the value used for the charm quark mass in the calculation, 
as shown in Fig.~\ref{fig:DzeroMNRScaling} (right) using the {\sf MNR} NLO calculation.
The scaling has a large $p_{\rm t}$ dependence in the low $p_{\rm t}$ region, where it varies from a factor of $\approx0.8$ at $p_{\rm t} \approx 1$~GeV/$c$ to $\approx0.4$ at $p_{\rm t} \approx 5$~GeV/$c$, while at higher $p_{\rm t}$ the variation less pronounced. 
The average scaling factor calculated for the \dzero, \dplus, and \dstar mesons is very similar, while some small variations can be observed on the uncertainty bands.

%
\subsubsection{Influence of the theoretical calculation: {\sf MNR} and {\sf GM-VFNS} vs {\sf FONLL}}

The {\sf FONLL} direct charm scaling has been tested by comparing the calculation to the one obtained with the {\sf MNR}~\cite{MNR} and {\sf GM-VFNS}~\cite{GMVFNS} calculations.

%
\paragraph{{\sf MNR} calculation:}

\begin{figure}[!htbp]
\begin{center}
        \includegraphics[width=0.48\textwidth]{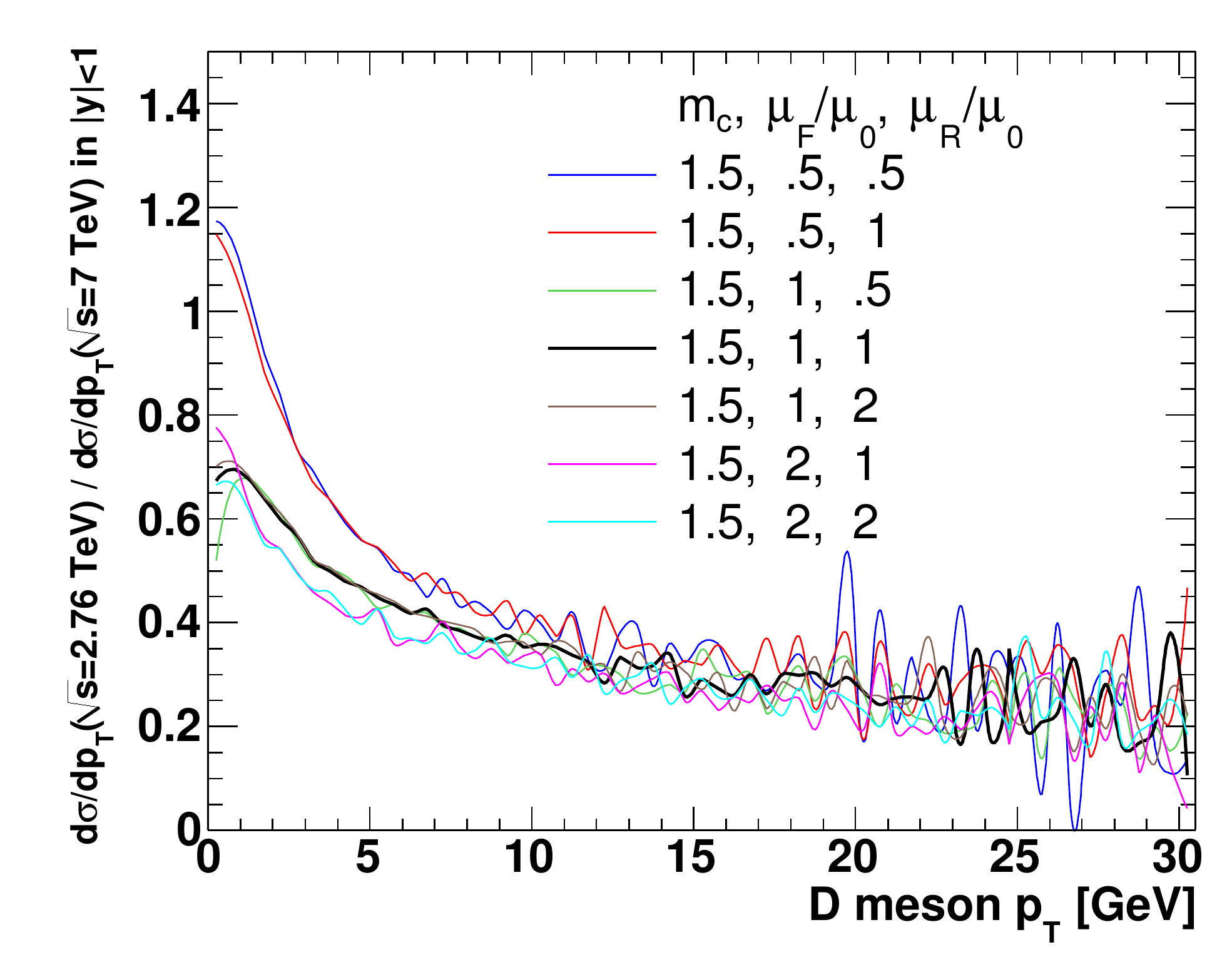}
        \includegraphics[width=0.46\textwidth]{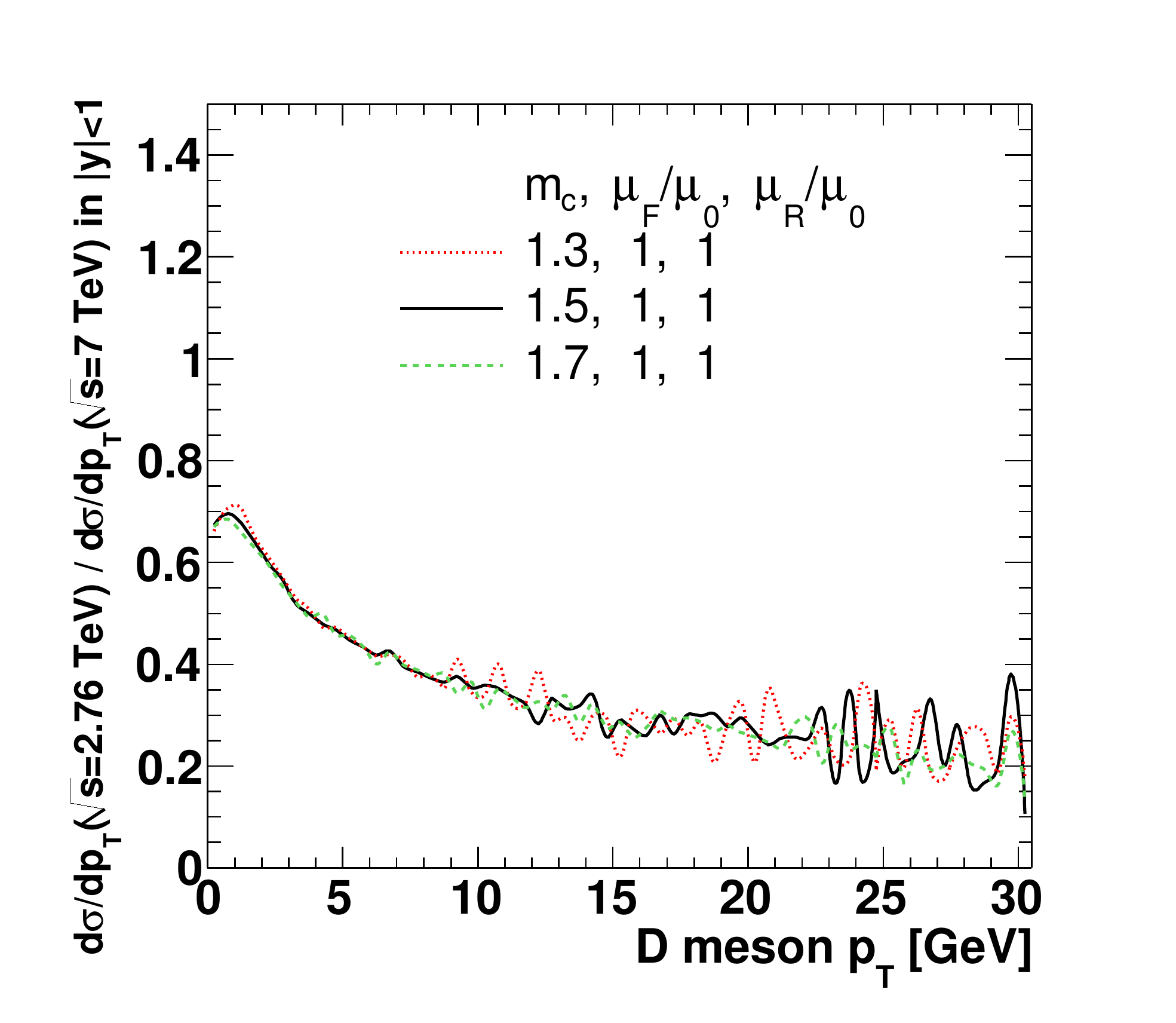}
\caption{\dzero~{\sf MNR} scaling to 2.76~TeV from 7 TeV (fine binning):
scales dependence (left) and charm quark mass dependence (right).}
\label{fig:DzeroMNRScaling}
\end{center}
\end{figure}

\begin{figure}[!htbp]
\begin{center}
        \includegraphics[width=.55\columnwidth]{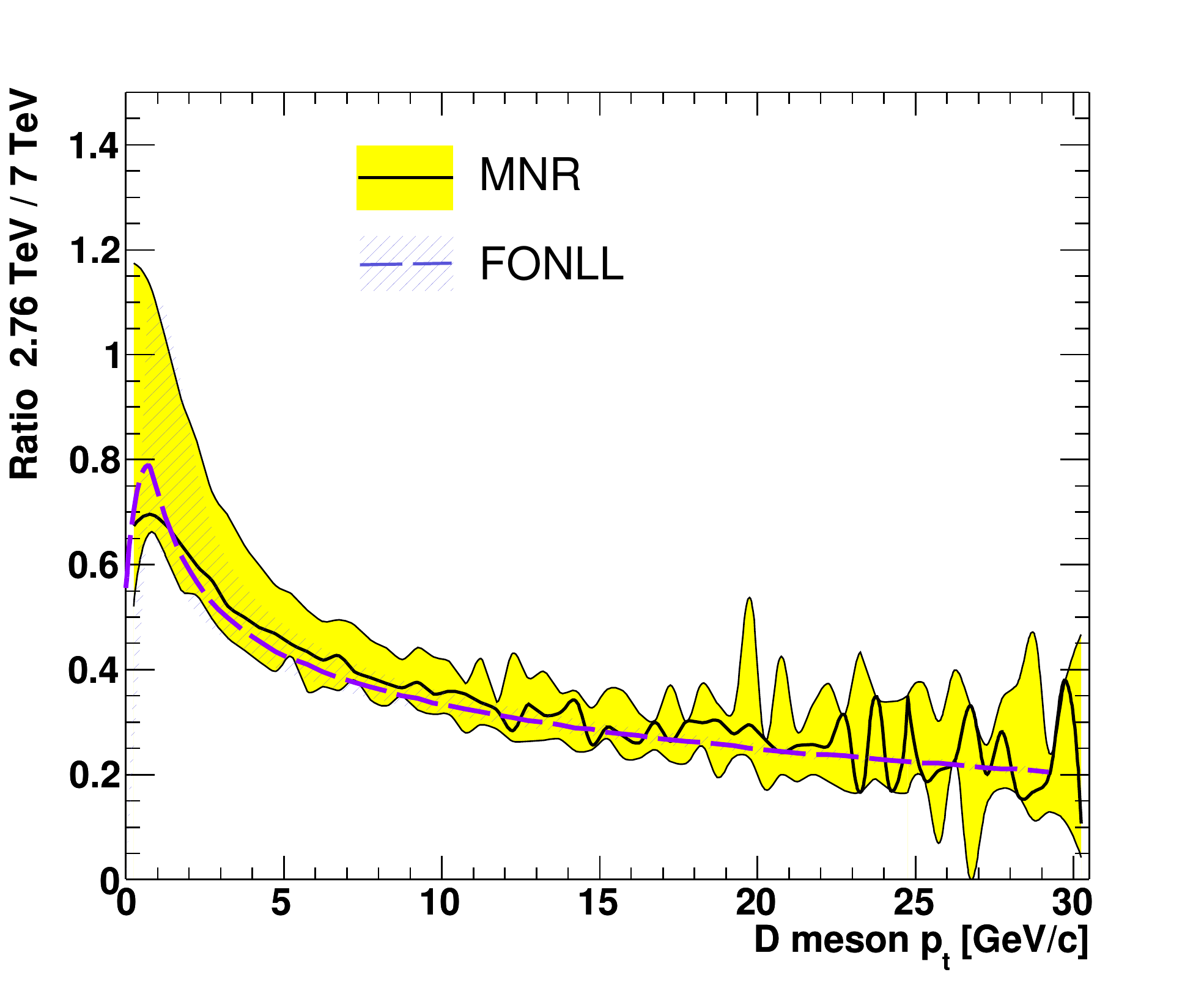}
\caption{\dzero~{\sf MNR} scaling to 2.76~TeV from 7 TeV (yellow band) compared to the {\sf FONLL} scaling (purple hatched band). }
\label{fig:DzeroMNRFONLLScaling}
\end{center}
\end{figure}

The {\sf MNR} scaling factor for the different sets of scales is shown in Fig.~\ref{fig:DzeroMNRScaling}~(left). Fig.~\ref{fig:DzeroMNRScaling}~(right) shows that the influence of varying the charm quark mass from 1.3 to 1.7 is negligible. 
The comparison of the {\sf MNR} and {\sf FONLL} calculations for \dzero~mesons (Fig.~\ref{fig:DzeroMNRFONLLScaling}) demonstrates that, as expected, the scaling factors agree with each other, and that the uncertainties are larger for the {\sf MNR} case. 
Therefore, from now on we will drop the {\sf MNR} case for this exercise. 

%
\paragraph{{\sf GM-VFNS} calculation:} 

\begin{figure}[!htbpt]
\begin{center}
        \includegraphics[width=0.59\columnwidth]{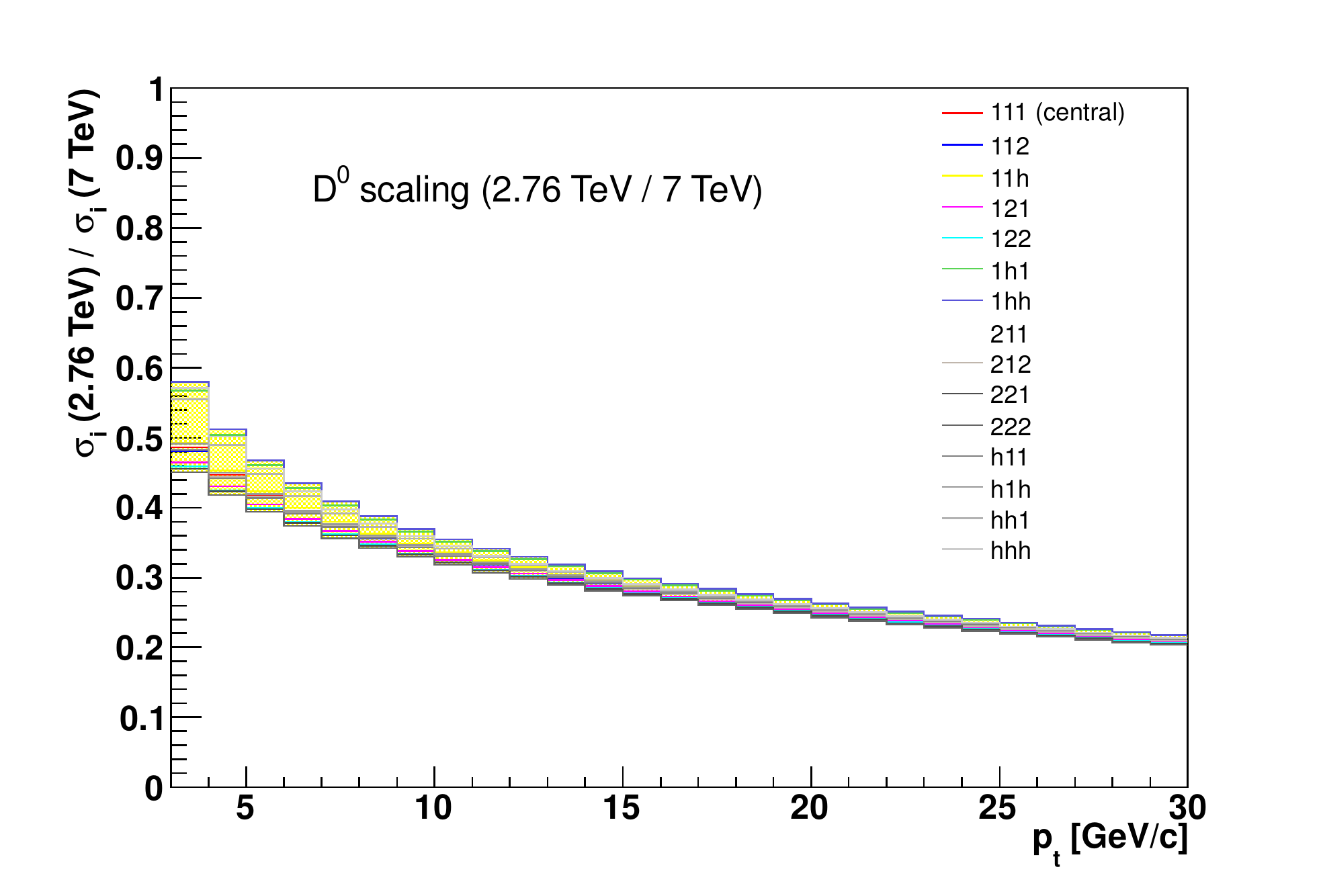}
\caption{\dzero~{\sf GM-VFNS} scaling to 2.76~TeV from 7 TeV considering that the scales are correlated vs energy. The yellow band represents the global scaling and its uncertainty. The values of the renormalization scale, the factorization scale for initial state singularities and the factorization scale for final state singularities are reported in the legend.
}
\label{fig:D0VFNSScaling}
\end{center}
\end{figure}

We obtain the {\sf GM-VFNS} scaling factor considering that the three calculation parameters (the renormalization scale, the factorization scale for initial state singularities and the factorization scale for final state singularities) do not depend on the value of $\sqrt{s}$, as for the {\sl FONLL} case,
with the difference that the latter considers only the factorization and renormalization scales. 
The \dzero~meson scaling to 2.76~TeV is shown in Fig.~\ref{fig:D0VFNSScaling}, where the calculation parameters are varied within $1/2$~(h), $1$ and $2$ times the standard parameters. 
The spread of the ratio evaluated for the different parameters indicate the scaling uncertainties. 
\begin{figure}[!htbp]
\begin{center}
        \includegraphics[width=0.495\columnwidth]{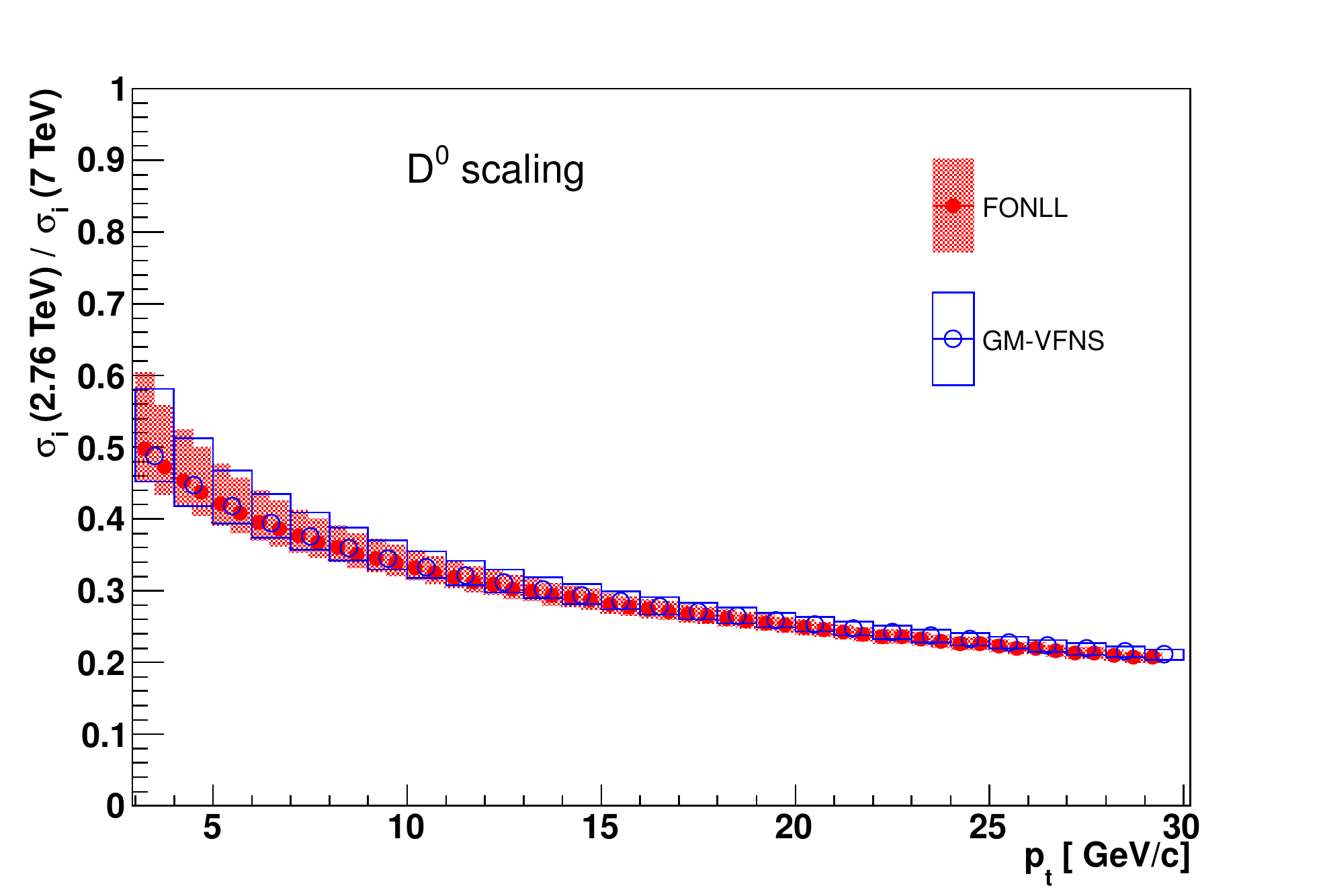}
        \includegraphics[width=0.495\columnwidth]{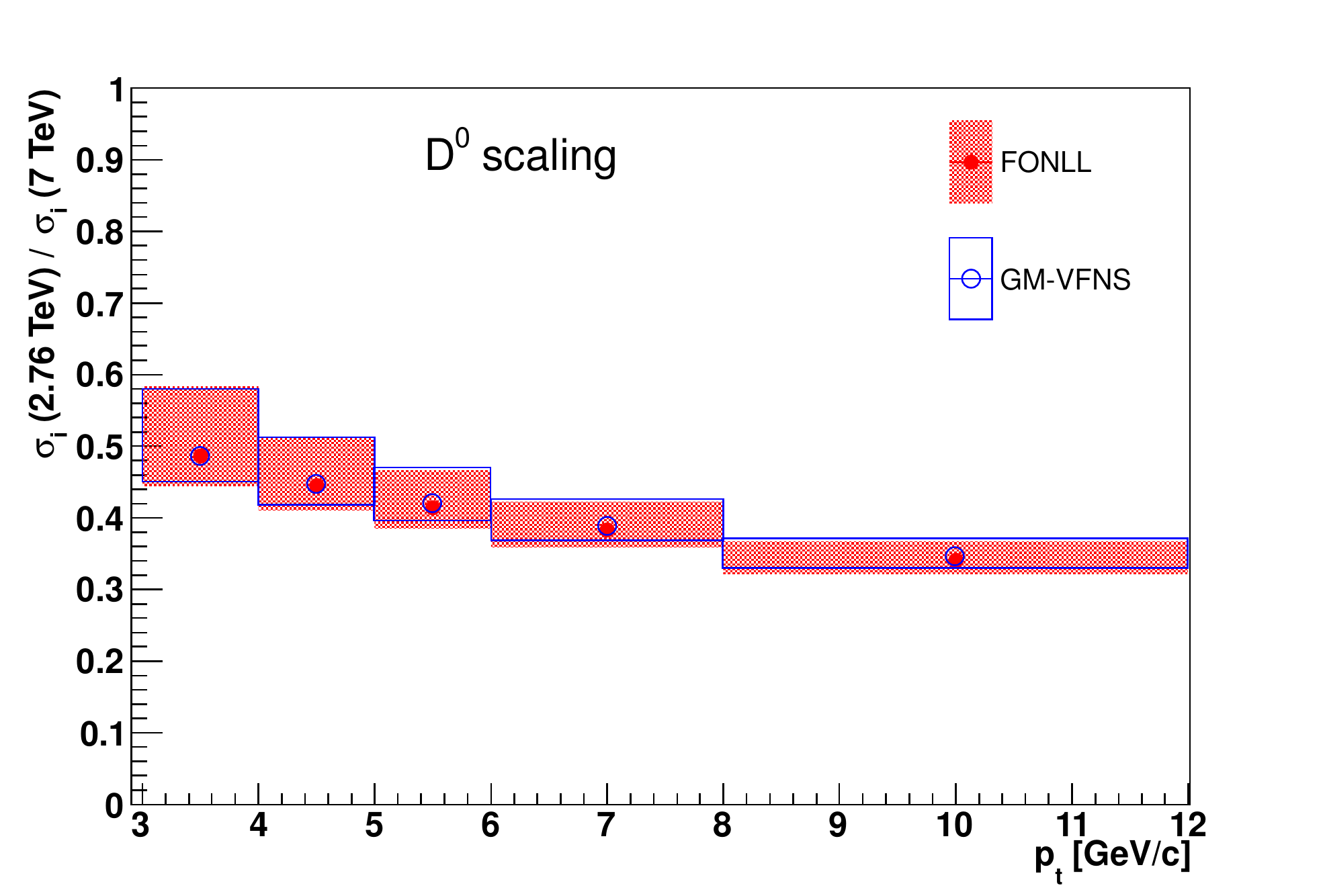}
\caption{Comparison of \dzero~{\sf GM-VFNS} and {\sf FONLL} scaling to 2.76~TeV from 7 TeV with fine~(left) and 7~TeV preliminary cross section binning (right). }
\label{fig:D0FONLLvsVFNSScaling}
\end{center}
\end{figure}

The comparison of the \dzero~{\sf FONLL} and {\sf GM-VFNS} scalings is shown in Fig.~\ref{fig:D0FONLLvsVFNSScaling} for different $p_{\rm t}$ binnings. The agreement of the scaling central values and their uncertainties for the considered $p_{\rm t}$ bins is striking. We can then conclude that there is no need to do all the scalings both with {\sf GM-VFNS} and {\sf FONLL} calculations since their energy evolution (and uncertainties) are in agreement. 
%

%
\subsubsection{Comparison to CDF measurements in $\rm p\overline p$ at $\sqrt{s}=1.96$~TeV}

\begin{figure}[!htbp]
\begin{center}
        \includegraphics[width=0.465\columnwidth]{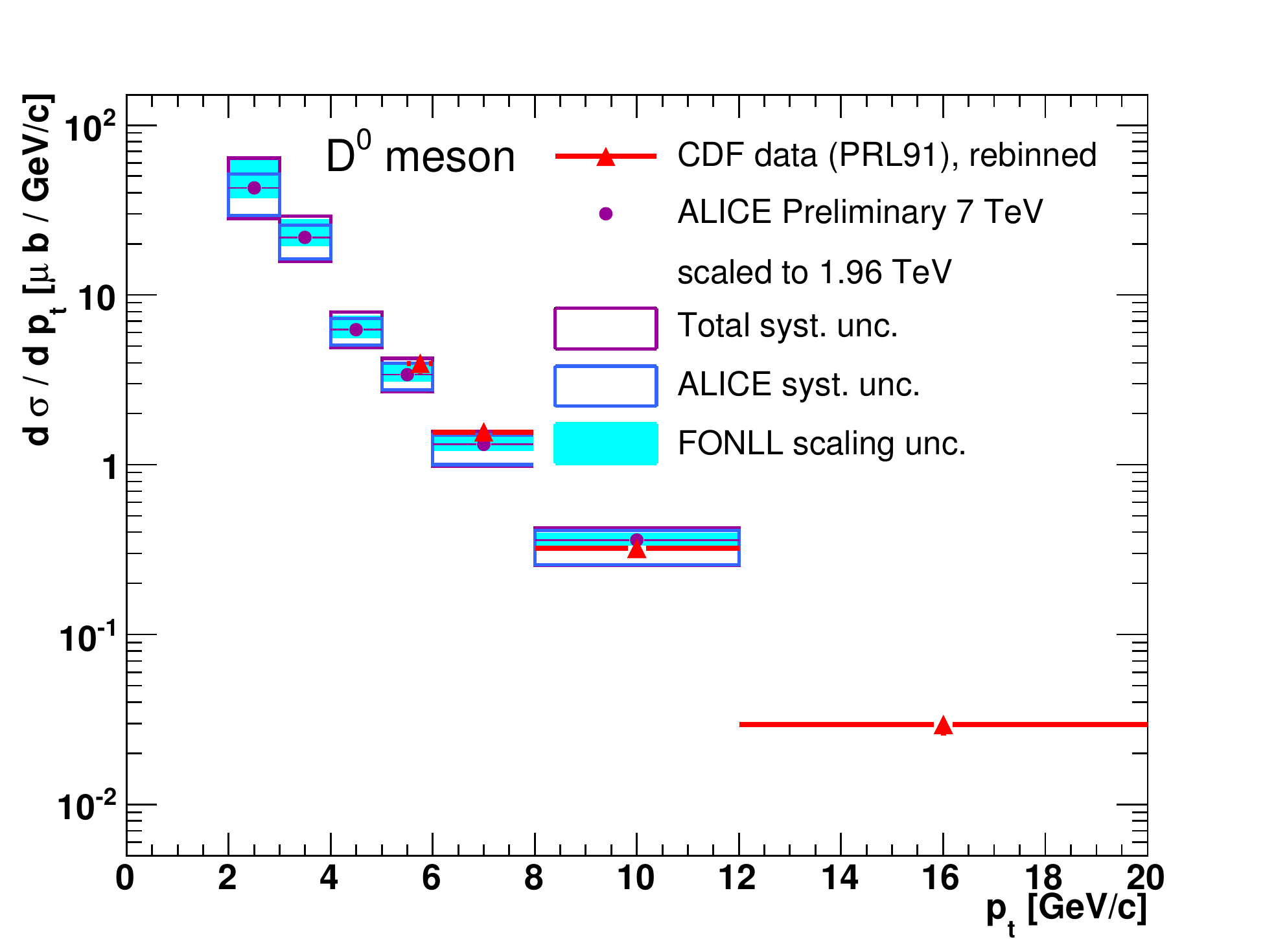}
        \includegraphics[width=0.515\columnwidth]{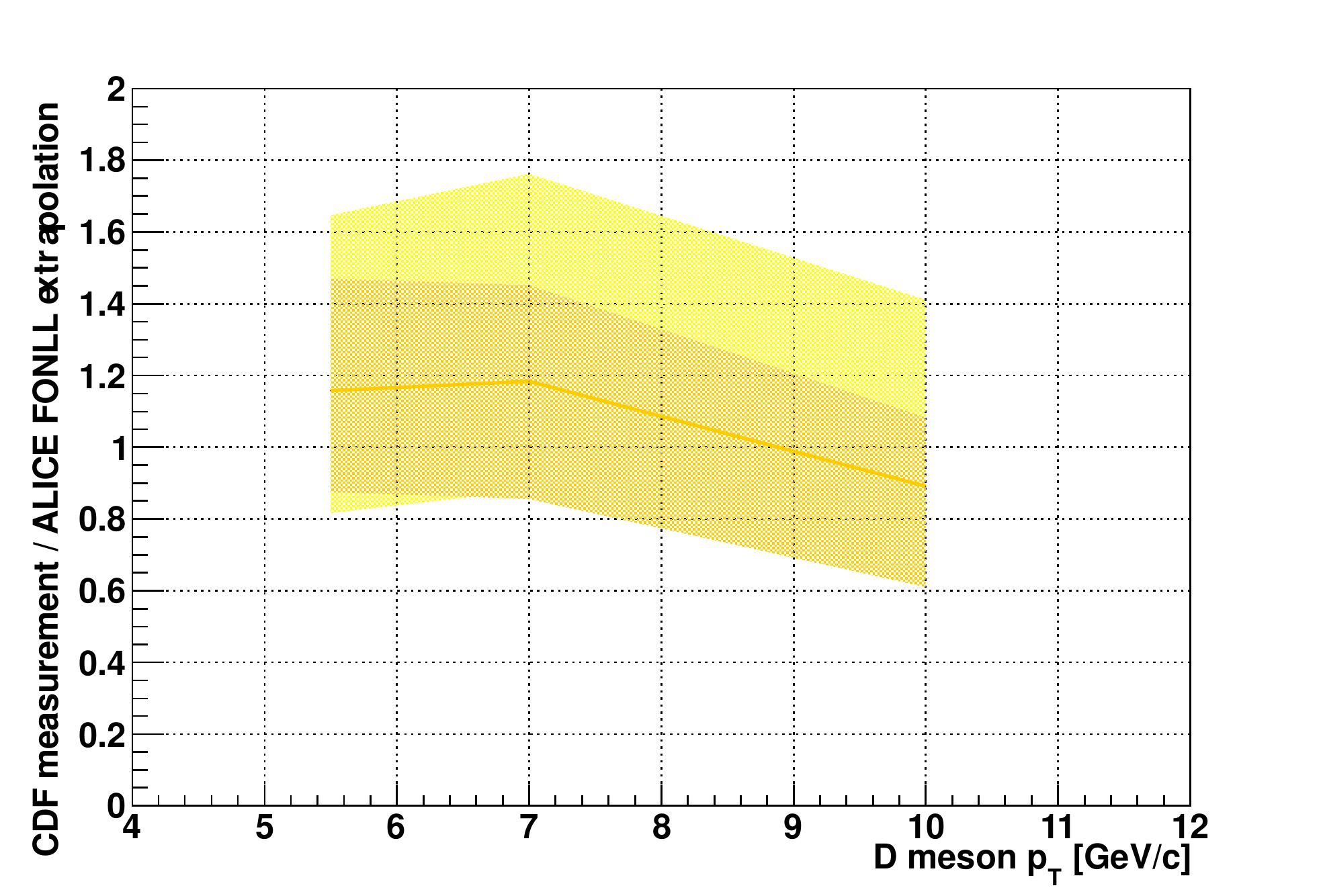}
\caption{Left: Comparison of \dzero~ALICE 7~TeV measurements scaled to 1.96~TeV with the CDF measurements. Right: ratio of these two. The yellow~(orange) band describes the maximum~(conservative) uncertainty on the ratio, considering that the CDF and ALICE scaled uncertainties are uncorrelated~(correlated). Note that the first CDF data point, $5.5<p_{\rm t}<6$~GeV/$c$, is compared to the ALICE data point for $5<p_{\rm t}<6$~GeV/$c$.}
\label{fig:D0ALICECDF_196TeV_newsigma}
\end{center}
\begin{center}
        \includegraphics[width=0.465\columnwidth]{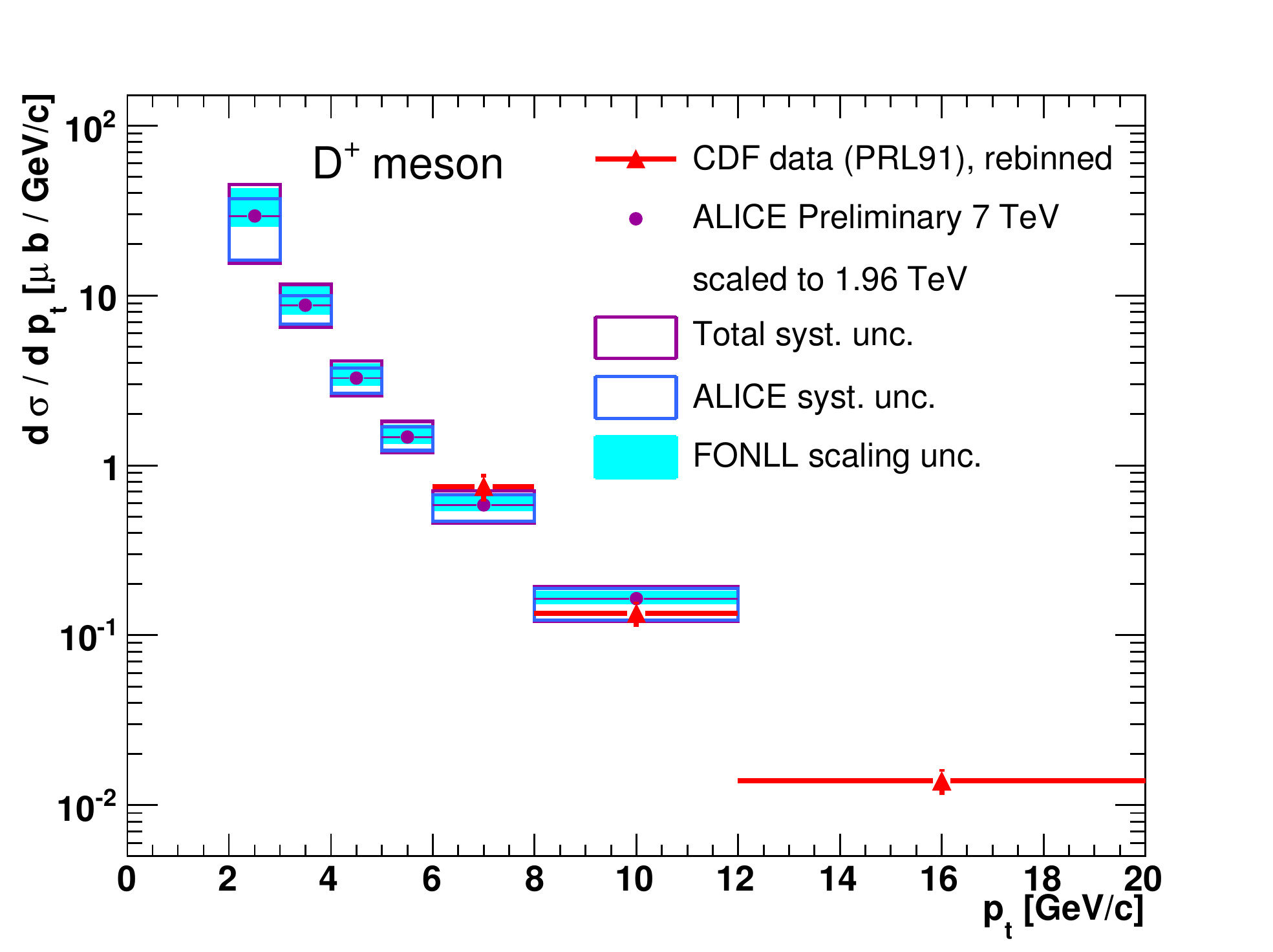}
        \includegraphics[width=0.515\columnwidth]{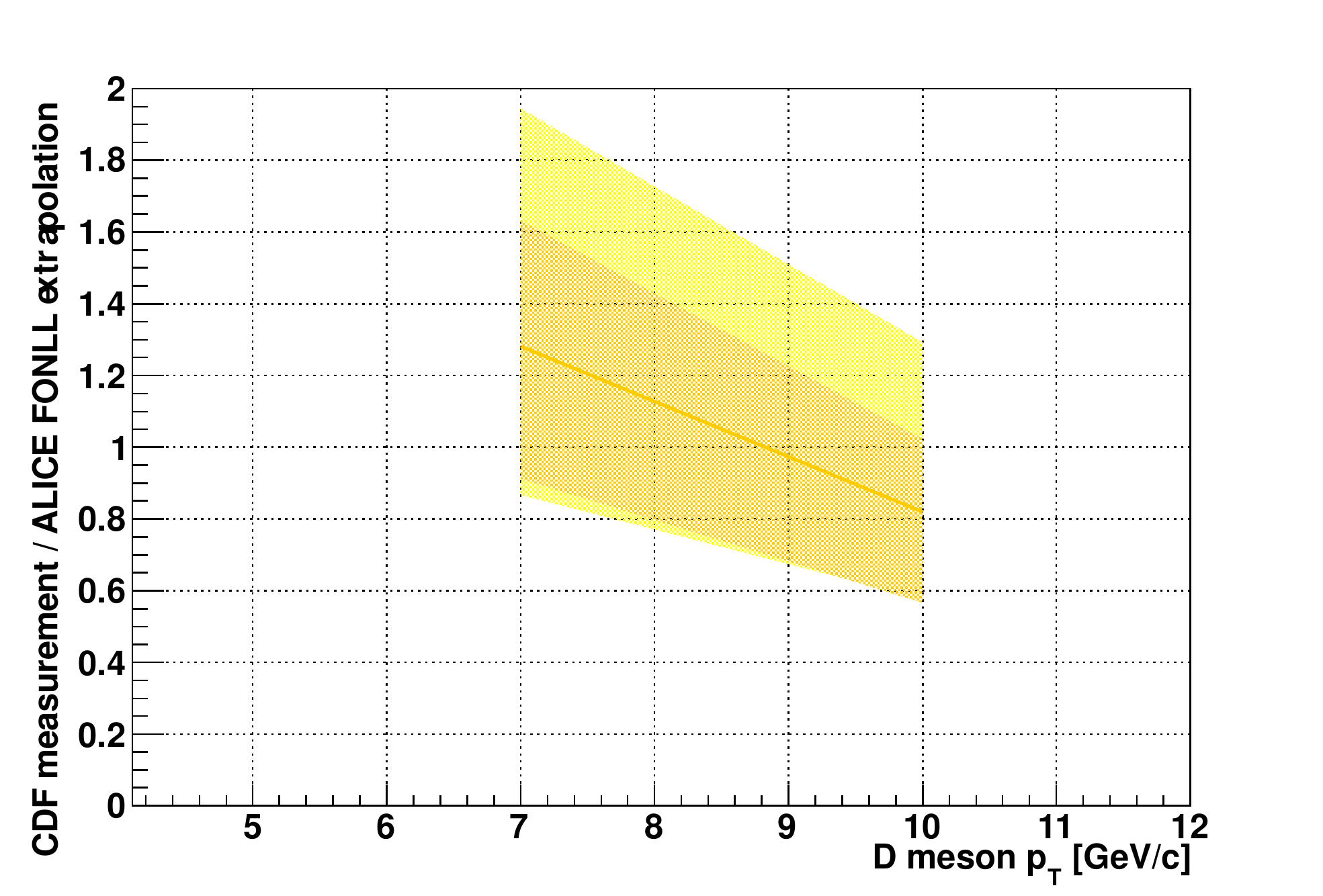}
\caption{Same as in Fig.~\ref{fig:D0ALICECDF_196TeV_newsigma}, for $D^{+}$.}
\label{fig:DplusALICECDF_196TeV_newsigma}
\end{center}
\begin{center}
        \includegraphics[width=0.465\columnwidth]{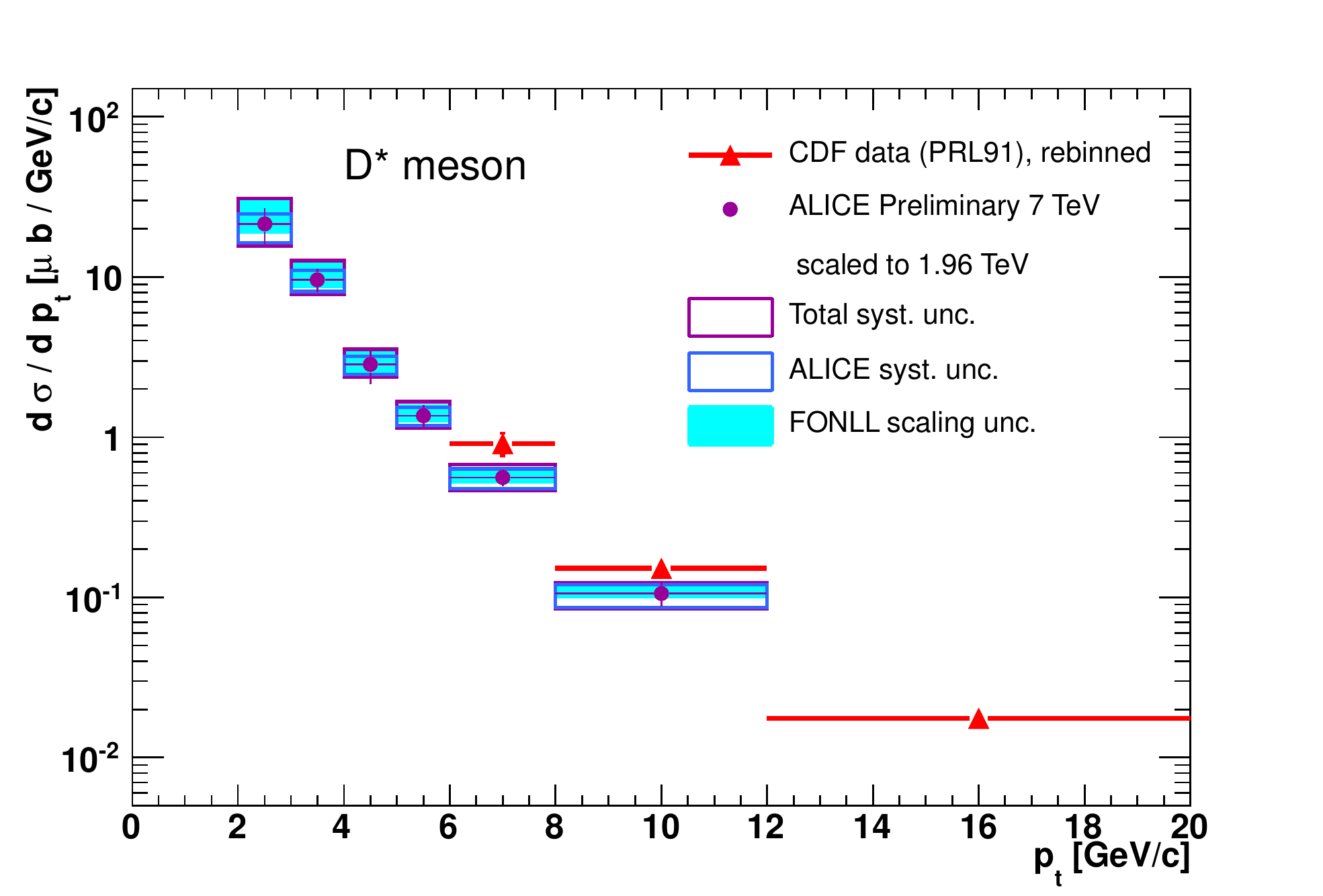}
        \includegraphics[width=0.515\columnwidth]{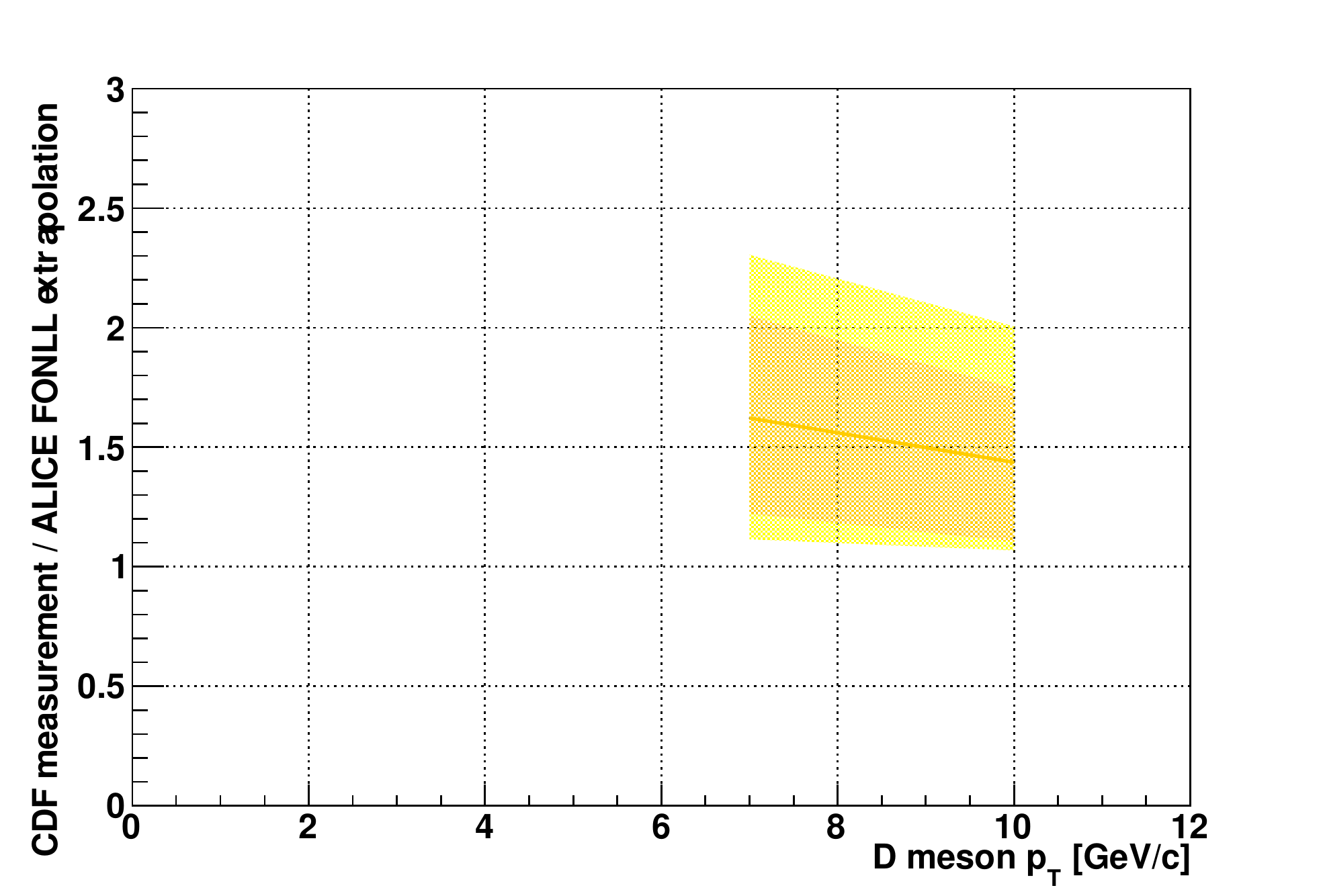}
\caption{Same as in Fig.~\ref{fig:D0ALICECDF_196TeV_newsigma}, for $D^{*+}$.}
\label{fig:DstarALICECDF_196TeV_newsigma}
\end{center}
\end{figure}

In this section, we show the comparison of the CDF~\cite{CDFdata} and the ALICE measurements scaled to 1.96~TeV.

We evaluated the scaling factor from 7~TeV to 1.96~TeV with the {\sf FONLL} calculations. These estimates were used to scale the \dzero, \dplus, and $D^{*+}$~cross sections measured by ALICE to 1.96~TeV. 
Figures~\ref{fig:D0ALICECDF_196TeV_newsigma}, \ref{fig:DplusALICECDF_196TeV_newsigma},
and \ref{fig:DstarALICECDF_196TeV_newsigma} 
present the comparison of these scalings and the CDF measurements\footnote{
Here the CDF cross sections have been rebinned to match the ALICE 7~TeV preliminary measurements binning. 
}. The ratio of the CDF / ALICE points is also shown (right-hand panels). The yellow~(orange) bands describe the maximum~(conservative) uncertainty on this ratio, considering that the CDF and ALICE scaled uncertainties are uncorrelated~(correlated), i.e. considering the ratios of the upper-CDF to lower-ALICE (upper-CDF to upper-ALICE) and vice versa. 
Overall, these ratios are compatible with unity (within somewhat large uncertainties), demonstrating that the scaling procedure is reliable. We note that, for the $D^{*+}$ case,
although compatible with 1 within 1.2 sigma, the ratio is centred at 1.5: 
rather than to an anomaly of the scaling, which is practically the same for all $D$ mesons,
this could related the observation that ratio $D^0/D^{*+}$ measured by ALICE at 7~TeV
is larger than that measured by CDF at 1.96~TeV~\cite{renu}.

%
\subsubsection{Results at 2.76~TeV and relative uncertainties}

\begin{figure}[!htbp]
\begin{center}
       \includegraphics[width=0.49\columnwidth]{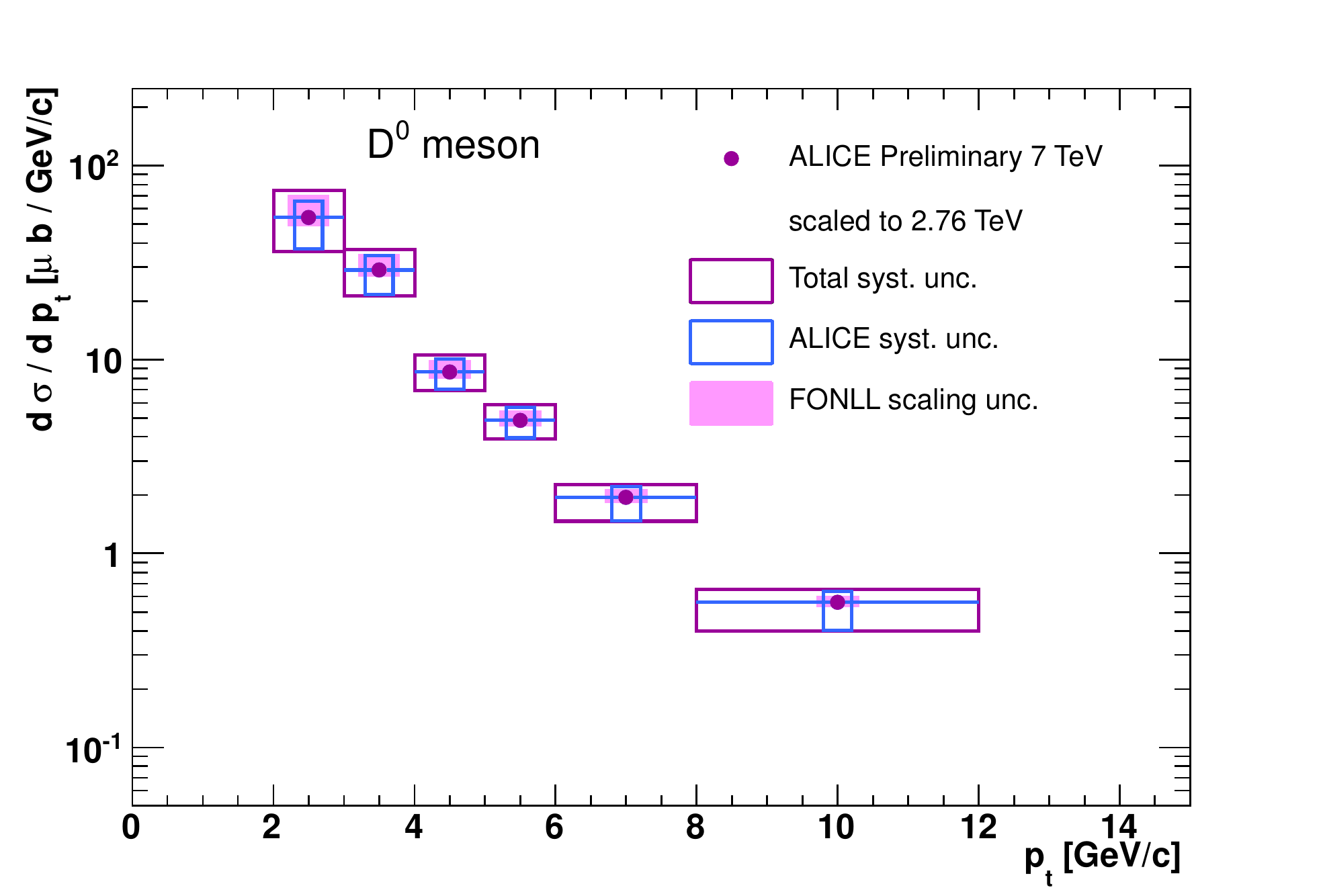}
        \includegraphics[width=0.49\columnwidth]{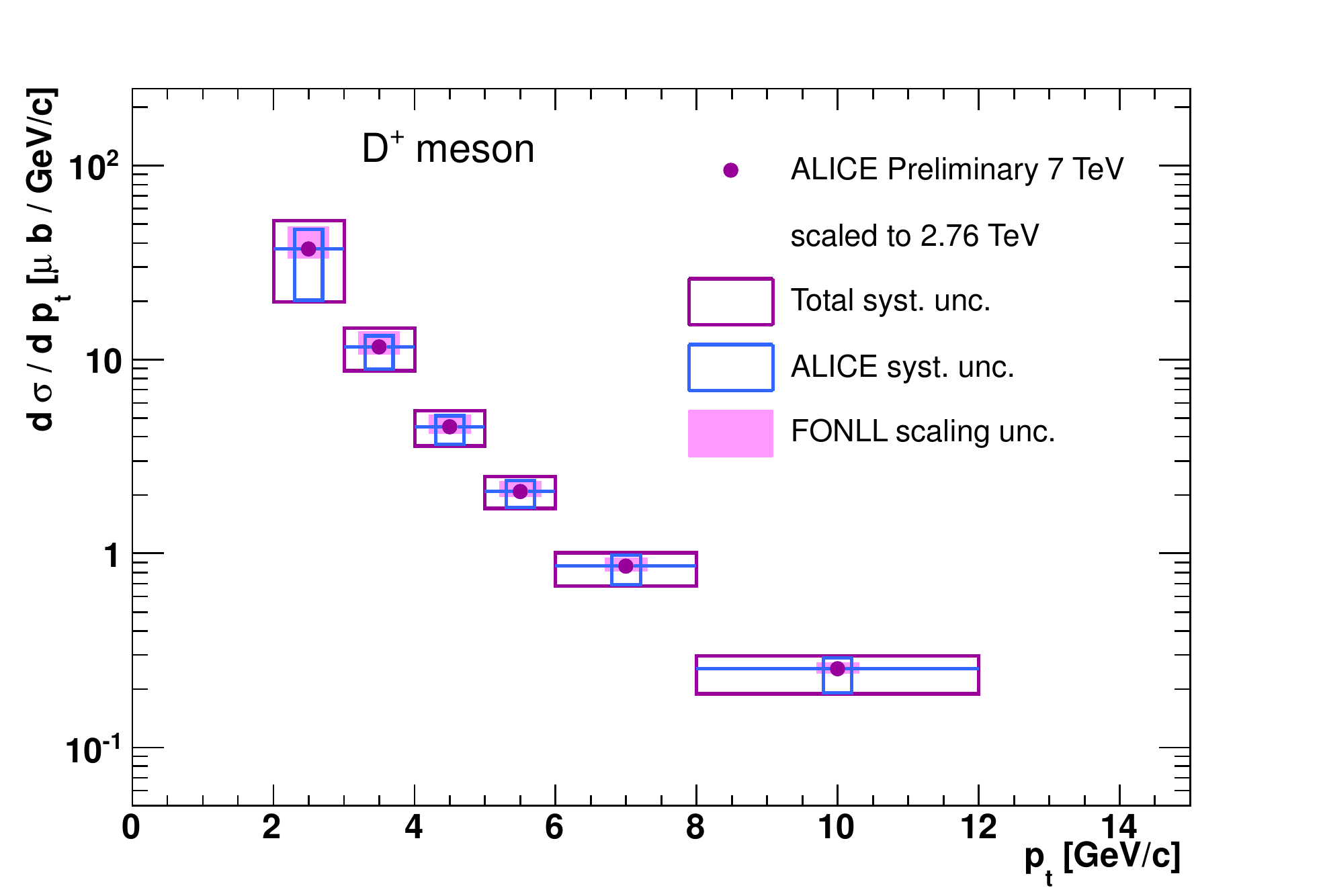}
        \includegraphics[width=0.49\columnwidth]{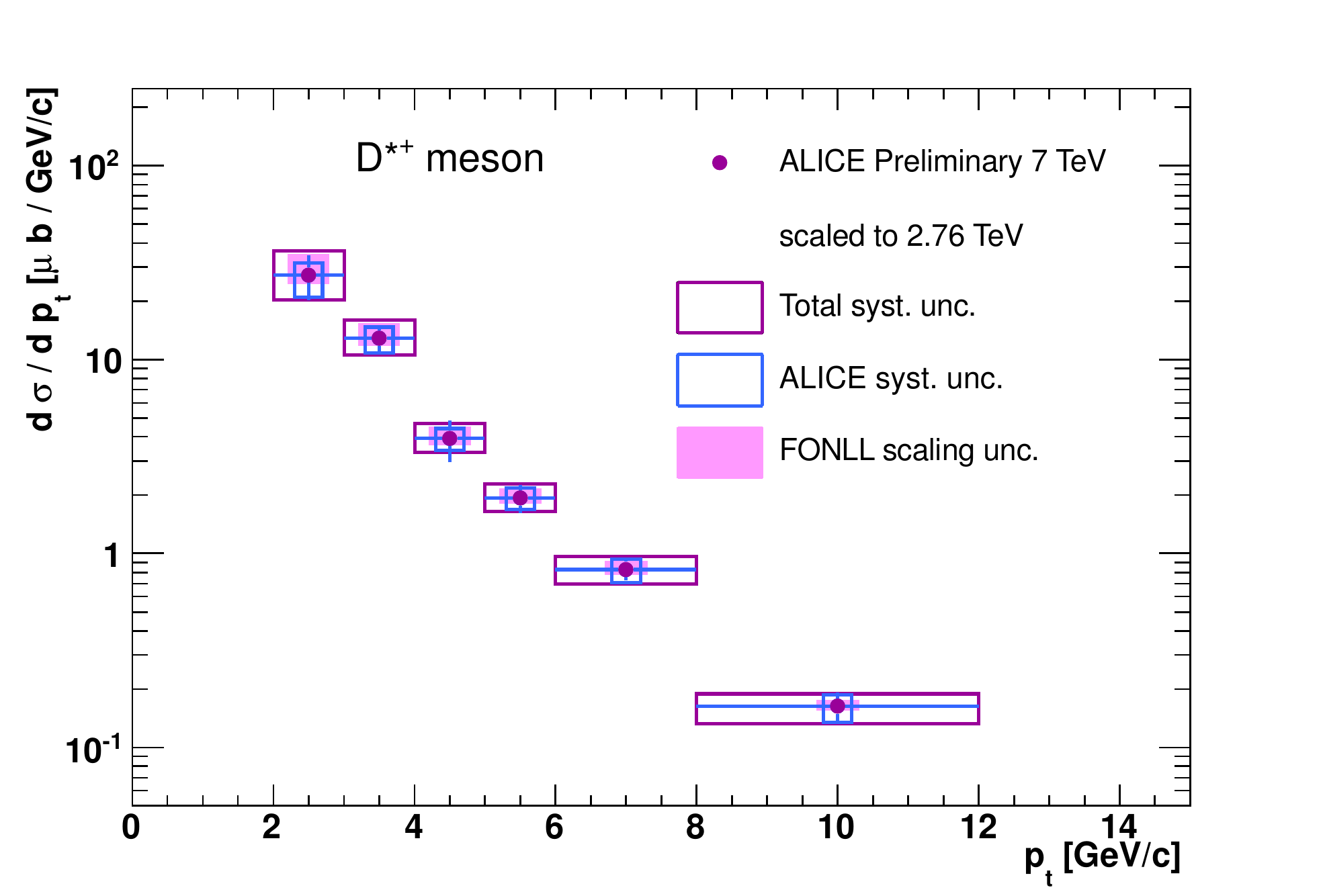}
\caption{\dzero~(top-left), \dplus~(top-right), and \dstar~(bottom) ALICE 7~TeV measurement scaling to 2.76~TeV. The purple points and boxes represent the interpolated cross section and its global systematics. The blue empty (magenta filled) boxes indicate the 7~TeV~measurement ({\sf FONLL} interpolation) contribution to the systematics.}
\label{fig:D0DplusDstarALICE_276TeV_newsigma}
\end{center}
\end{figure}

Finally, we can scale the ALICE \dzero, \dplus~and \dstar~measurements at~7~TeV to 2.76~TeV considering the {\sf FONLL} scaling factors evaluated in section~\ref{sec:CharmScaling276TeV}. The scaled cross sections are presented in Fig.~\ref{fig:D0DplusDstarALICE_276TeV_newsigma} for \dzero~(top-left), \dplus~(top-right) and \dstar~(bottom). The influence of the 7~TeV data systematics and of the {\sf FONLL} interpolation systematics on the global scaled systematics is also depicted, showing the relatively small contribution of the {\sf FONLL} scaling uncertainties.

\subsection{Heavy flavour decay electrons in $|y|<0.8$}
\label{sec:ElectronScaling}

The FONLL scaling factor from $\sqrt{s} = 7$~TeV to $\sqrt{s} = 2.76$~TeV
is calculated for electrons from charm and beauty decays using the approach
described in section~\ref{sec:recipe}. For all charm related calculations 
shown in this section we assume that neutral $D$ mesons contribute 70\% to 
the total electron yield and the remaining 30\% originate from charged $D$ 
meson decays.

\begin{figure}[!htbp]
\begin{center}
  \includegraphics[width=0.475\columnwidth]{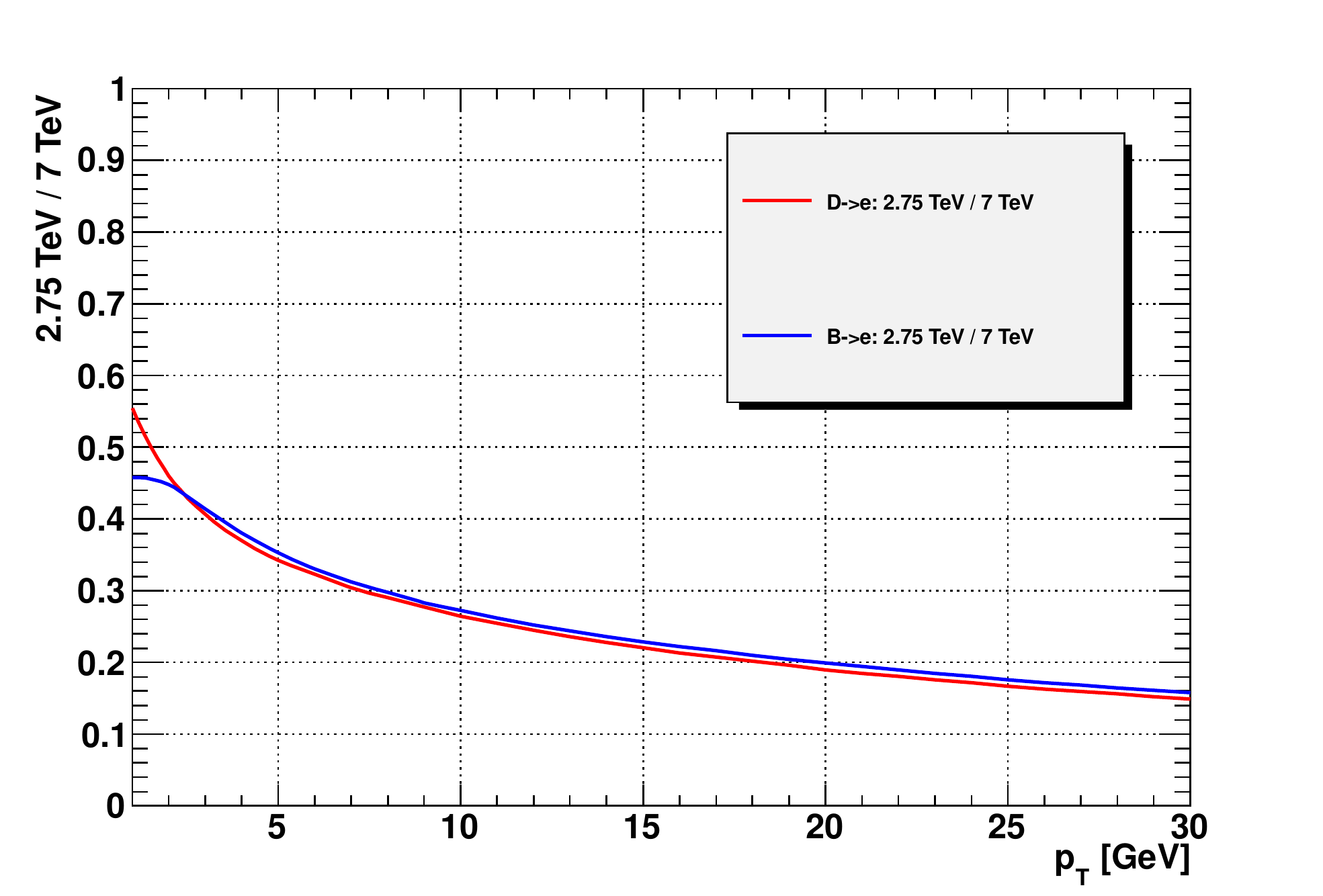}
  \includegraphics[width=0.475\columnwidth]{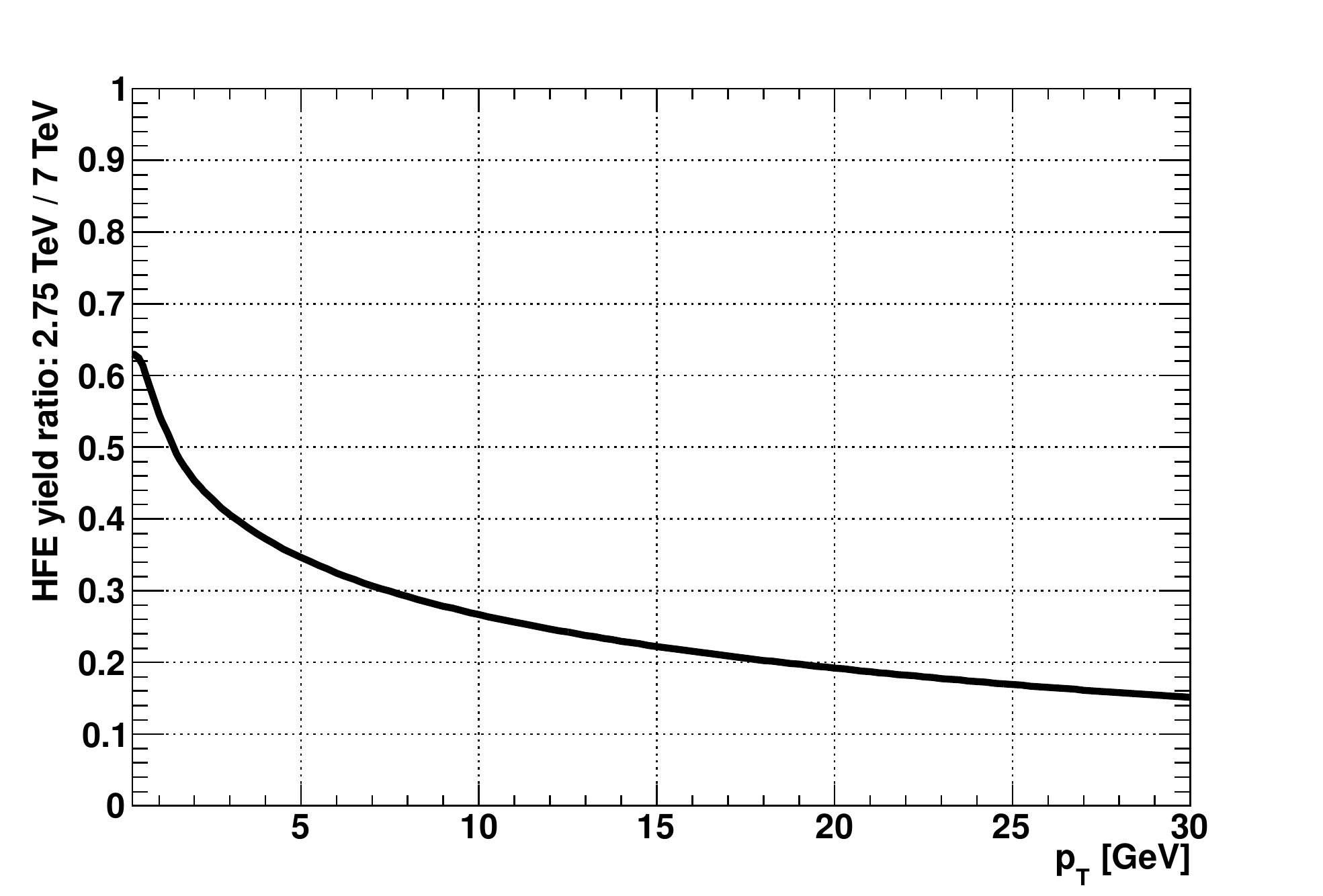}
\caption{FONLL scaling factor from 7 TeV to 2.76 TeV for electrons from charm
and beauty decays, respectively (left panel). Combined FONLL scaling factor for
electron from heavy flavour decays (right panel).}
\label{fig:ce_vs_be}
\end{center}
\end{figure}

In the case of electrons from heavy flavour decays an additional 
complication arises from the fact that the relative contributions from 
charm and beauty decays change as function of $p_{\rm t}$ and are not known 
a priori. Therefore, it is crucial to compare the scaling for electrons
from charm decays and beauty decays separately before evaluating a combined
scaling function. This comparison is shown in the left panel of
Fig.~\ref{fig:ce_vs_be} for the default choices of quark masses, parton
distribution function, and scales $\mu_R$ and $\mu_F$. The scaling factors 
for electrons from charm and beauty decays are almost the same except for 
the region of low transverse momenta ($p_{\rm t} < 2$~GeV/$c$), where the relative 
contribution from beauty decays to the total heavy flavour decay electron 
yield is tiny.

The agreement of the charm and beauty decay electron scaling factors as observed
in FONLL justifies to calculate one combined scaling factor for heavy flavour 
decay electrons as the ratio of the sum of charm and beauty decay electron 
cross sections at 2.76 TeV relative to the sum at 7 TeV. This combined scaling
factor is shown in the right panel of Fig.~\ref{fig:ce_vs_be} for the default
FONLL parameters.

In the following, we discuss the uncertainties of the combined heavy flavour
decay electron scaling factor due to the uncertainties of the FONLL parameters,
{\it i.e.} the uncertainties of the quark masses and the scale parameters
$\mu_R$ and $\mu_F$.

\begin{figure}[!htbp]
\begin{center}
  \includegraphics[width=0.465\columnwidth]{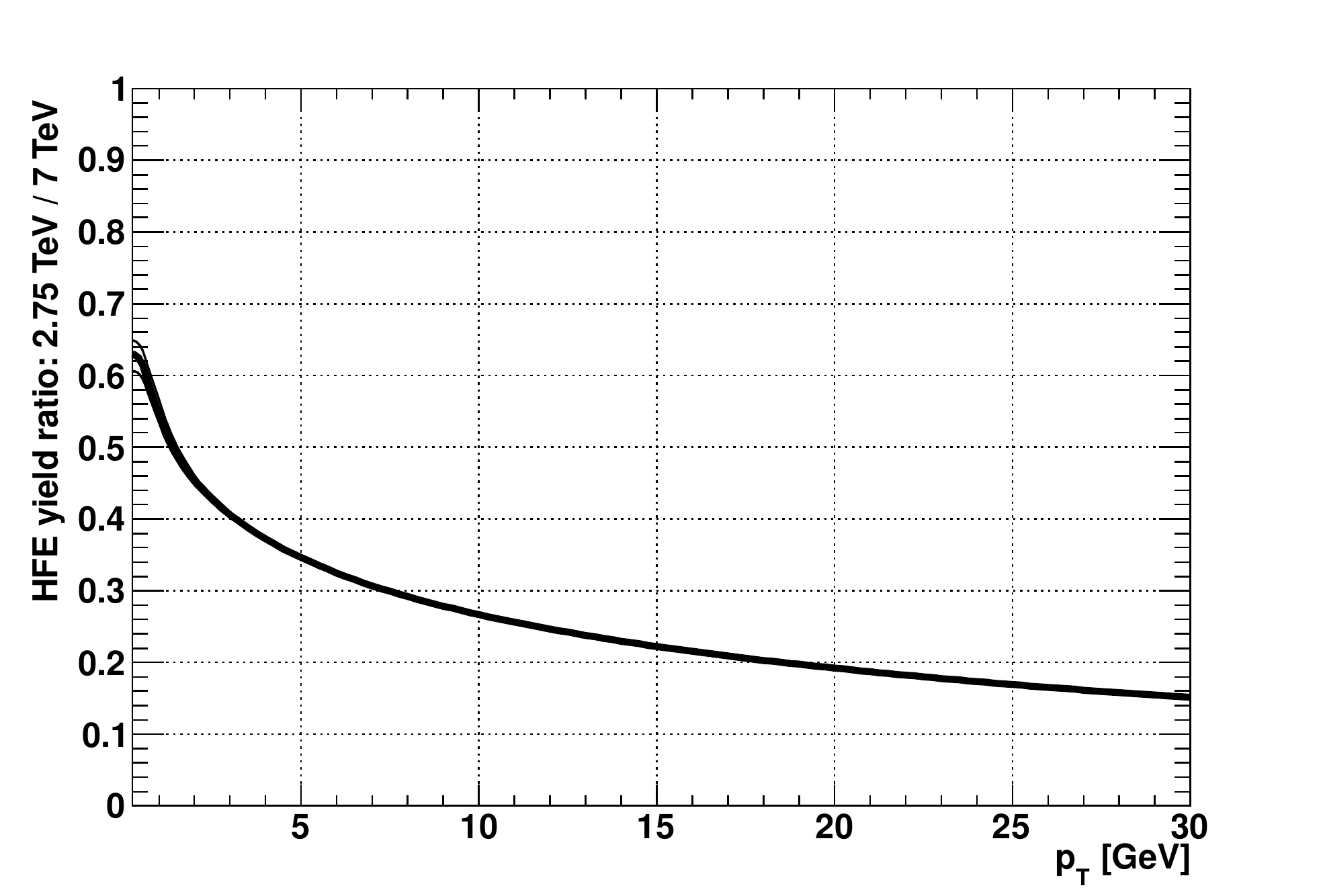}
  \includegraphics[width=0.465\columnwidth]{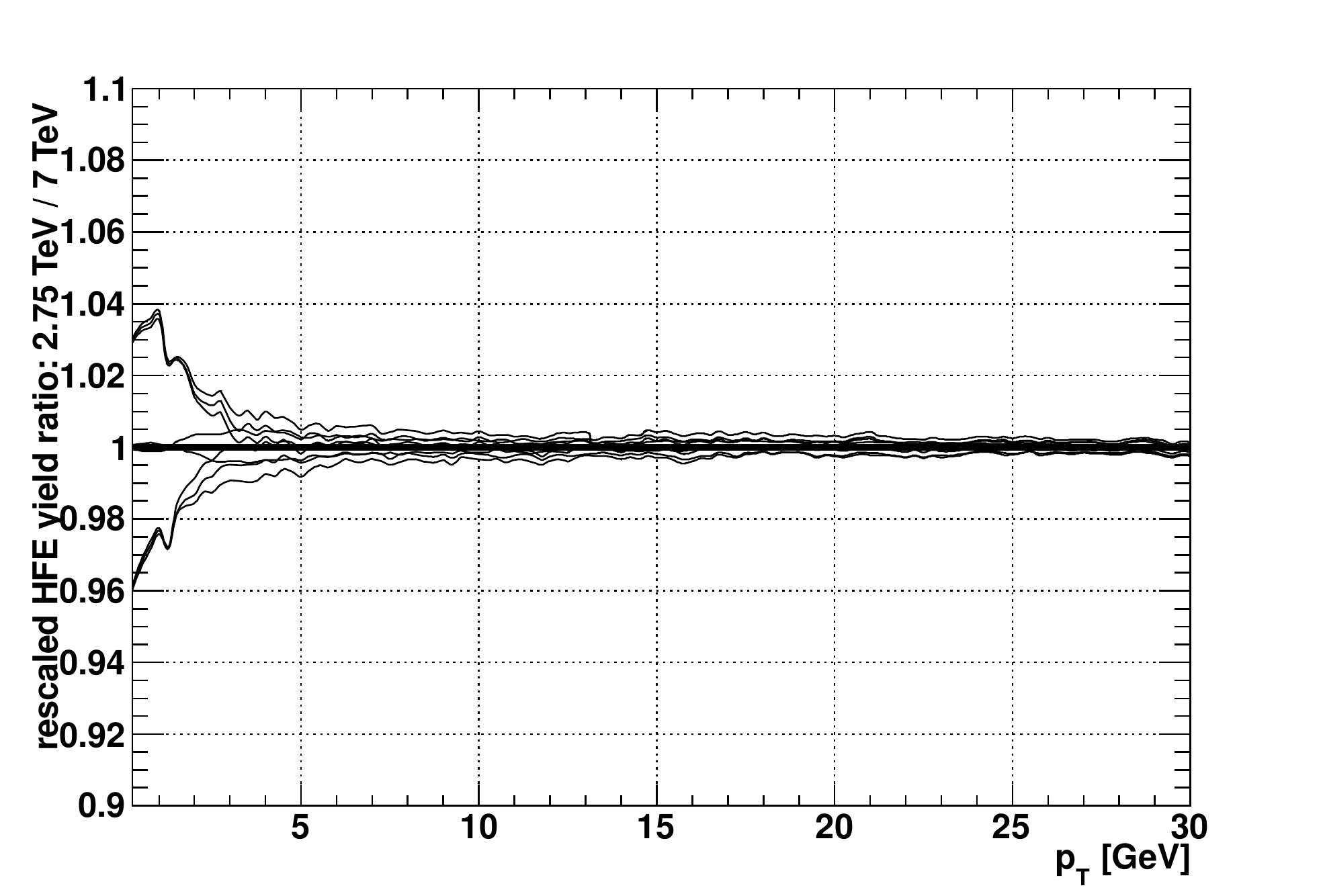}
\caption{FONLL scaling factor from 7 TeV to 2.76 TeV for electrons from heavy
flavour decays with different values for the bare quark masses (left panel). 
Variation of scaling factors obtained with different quark masses relative to 
the scaling factor calculated with the default quark masses (right panel).}
\label{fig:hfe_mass}

  \includegraphics[width=0.465\columnwidth]{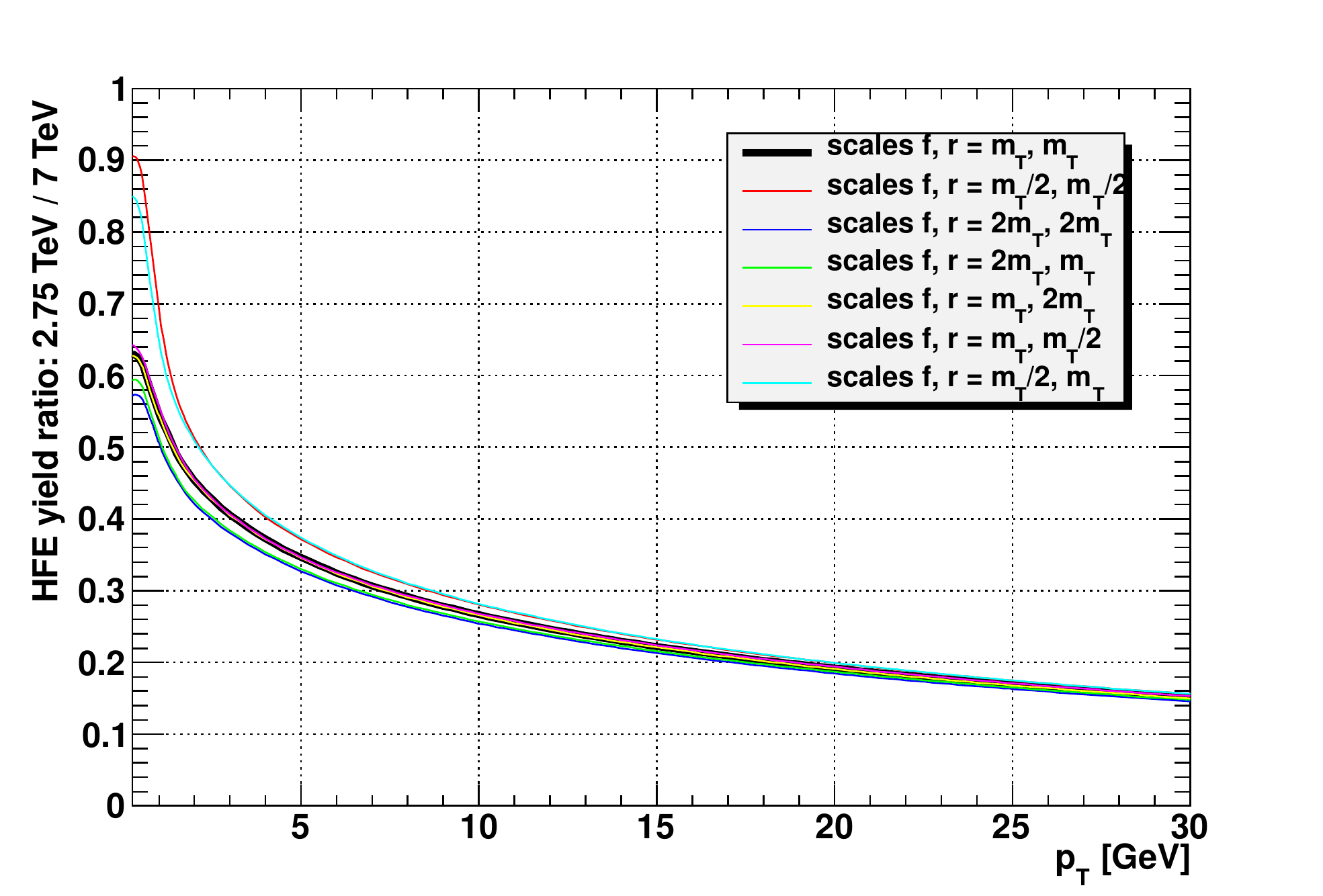}
  \includegraphics[width=0.465\columnwidth]{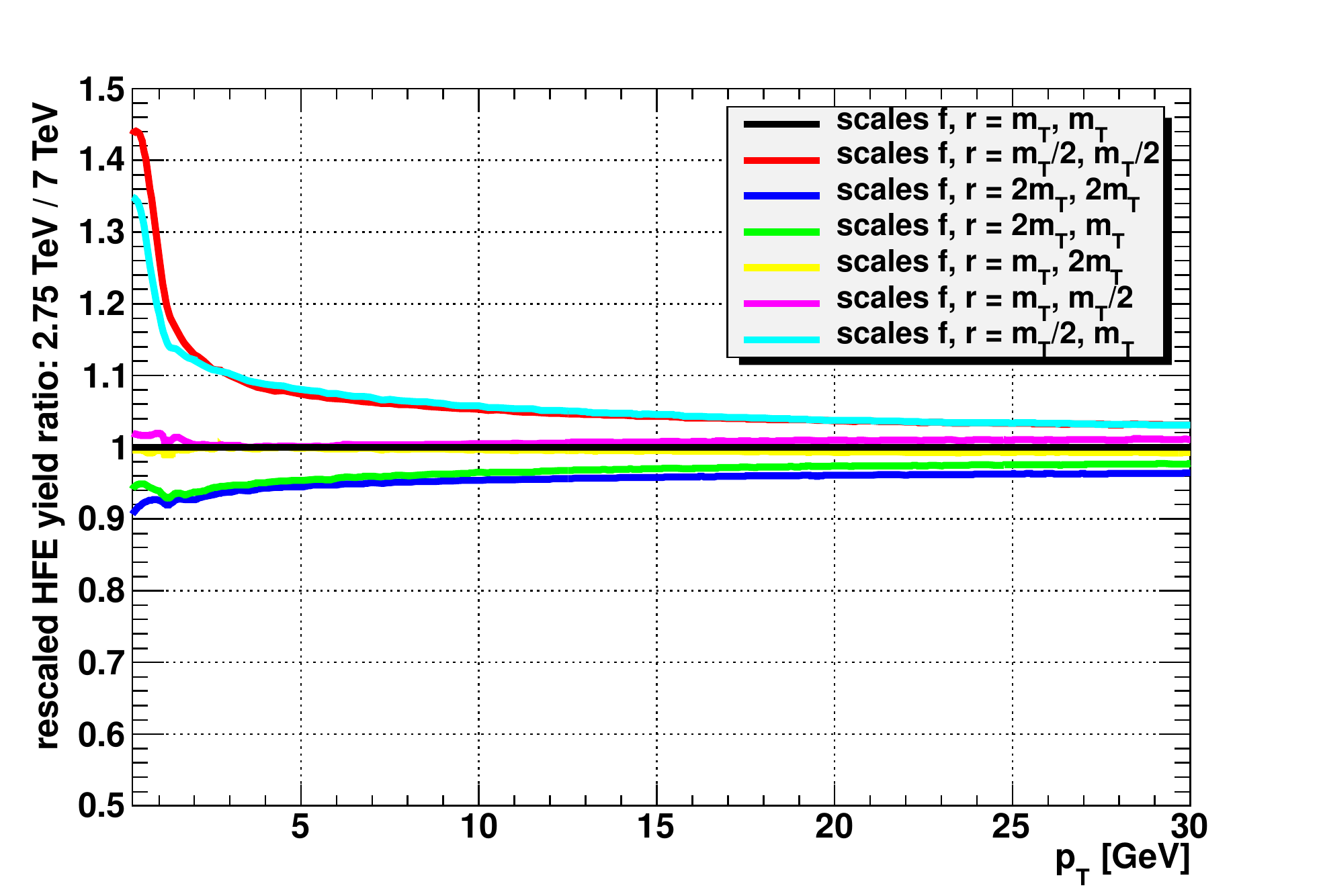}
\caption{FONLL scaling factor from 7 TeV to 2.76 TeV for electrons from heavy
flavour decays with different values for the scale parameters $\mu_R$ and 
$\mu_F$, which are chosen to be the same for charm and beauty (left panel). 
Variation of scaling factors obtained with different scale parameters relative 
to the scaling factor calculated with the default scale parameters (right 
panel).}
\label{fig:hfe_scale_same}

  \includegraphics[width=0.465\columnwidth]{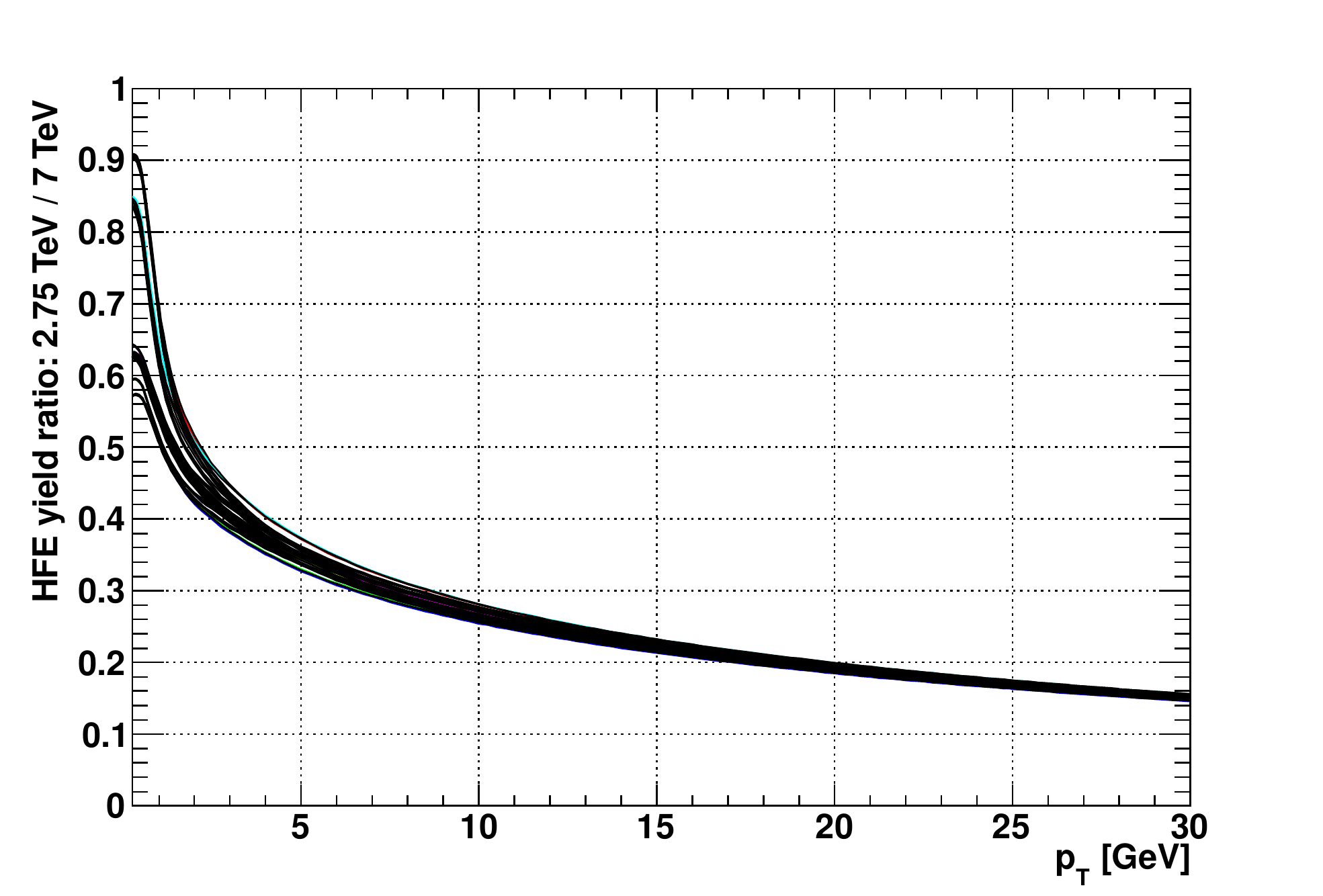}
  \includegraphics[width=0.465\columnwidth]{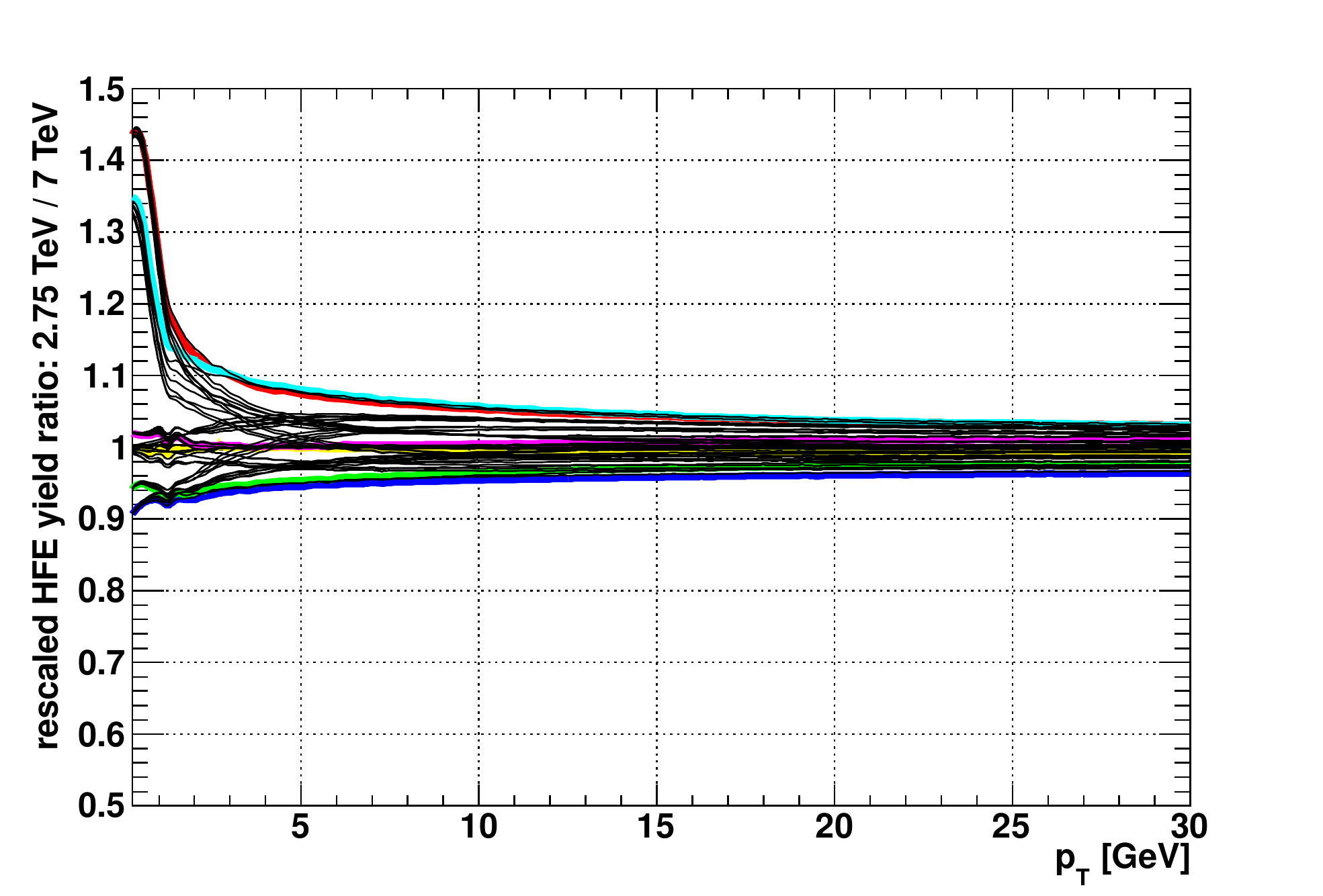}
\caption{FONLL scaling factor from 7 TeV to 2.76 TeV for electrons from heavy
flavour decays with different values for the scale parameters $\mu_R$ and 
$\mu_F$, where the scale parameters are allowed to vary independently for charm 
and beauty (left panel). Variation of scaling factors obtained with different 
scale parameters relative to the scaling factor calculated with the default 
scale parameters (right panel).}
\label{fig:hfe_scale_indep}
\end{center}
\end{figure}

The dependence of the scaling factor on the quark masses turns out to
be negligible as demonstrated in Fig.~\ref{fig:hfe_mass}. Here, we consider
quark masses of $m_c = 1.5 \pm 0.2$~GeV and $m_b = 4.75 \pm 0.25$~GeV.

As for $D$ mesons, 
the dependence of the heavy flavour decay electron scaling factor on the choice
of the FONLL scale parameters is addressed in Figs.~\ref{fig:hfe_scale_same}
and \ref{fig:hfe_scale_indep}. In addition to the FONLL default scales
$\mu_R = \mu_F = \mu_0$, we calculate the scaling factor for the scale values
$0.5\,\mu_0$ and $2\,\mu_0$.

First, we assume that the scales are the same for charm and beauty. The
resulting heavy flavour electron scaling factors are shown in 
Fig.~\ref{fig:hfe_scale_same}. The spread of the calculations with
different scaling factors is of order 10\% or less for an electron $p_{\rm t}$ 
larger than 2 GeV/$c$.

However, it can not be excluded that the scale parameters vary independently
for charm and beauty. To quantify the uncertainty due to that possibility
we show in Fig.~\ref{fig:hfe_scale_indep} the heavy flavour electron
scaling factors one can calculate with individual choices for the scale
parameters for charm and beauty in FONLL. The resulting spread of these
calculations is within the envelope of the calculations for which the
same scale parameters have been used for charm and beauty.  

\begin{figure}[!htbp]
\begin{center}
  \includegraphics[width=0.475\columnwidth]{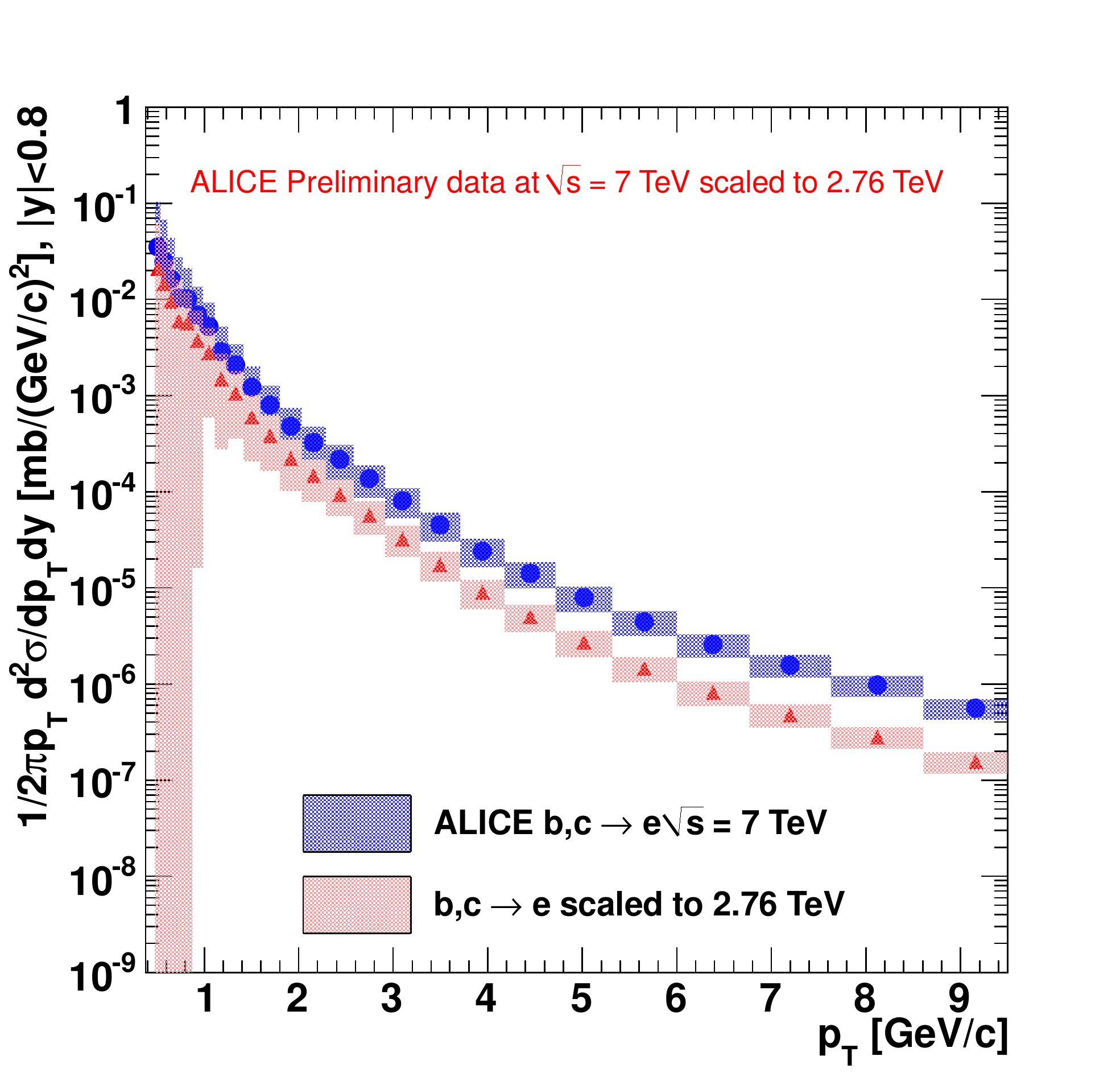}
\caption{Cross section of electrons from heavy flavour decays
measured by ALICE at $\sqrt{s}=7$~TeV~\cite{noteHFE} and the result of the scaling to 2.76~TeV.}
\label{fig:electronsscaled}
\end{center}
\end{figure}

Figure~\ref{fig:electronsscaled} shows the cross section of electrons from heavy flavour decays
measured by ALICE at $\sqrt{s}=7$~TeV and the result of the scaling to 2.76~TeV.

\subsection{Heavy flavour decay muons in $2.5<y<4$}
\label{sec:MuonScaling}

\begin{figure}[!htbp]
\begin{center}
        \includegraphics[width=0.479\columnwidth]
{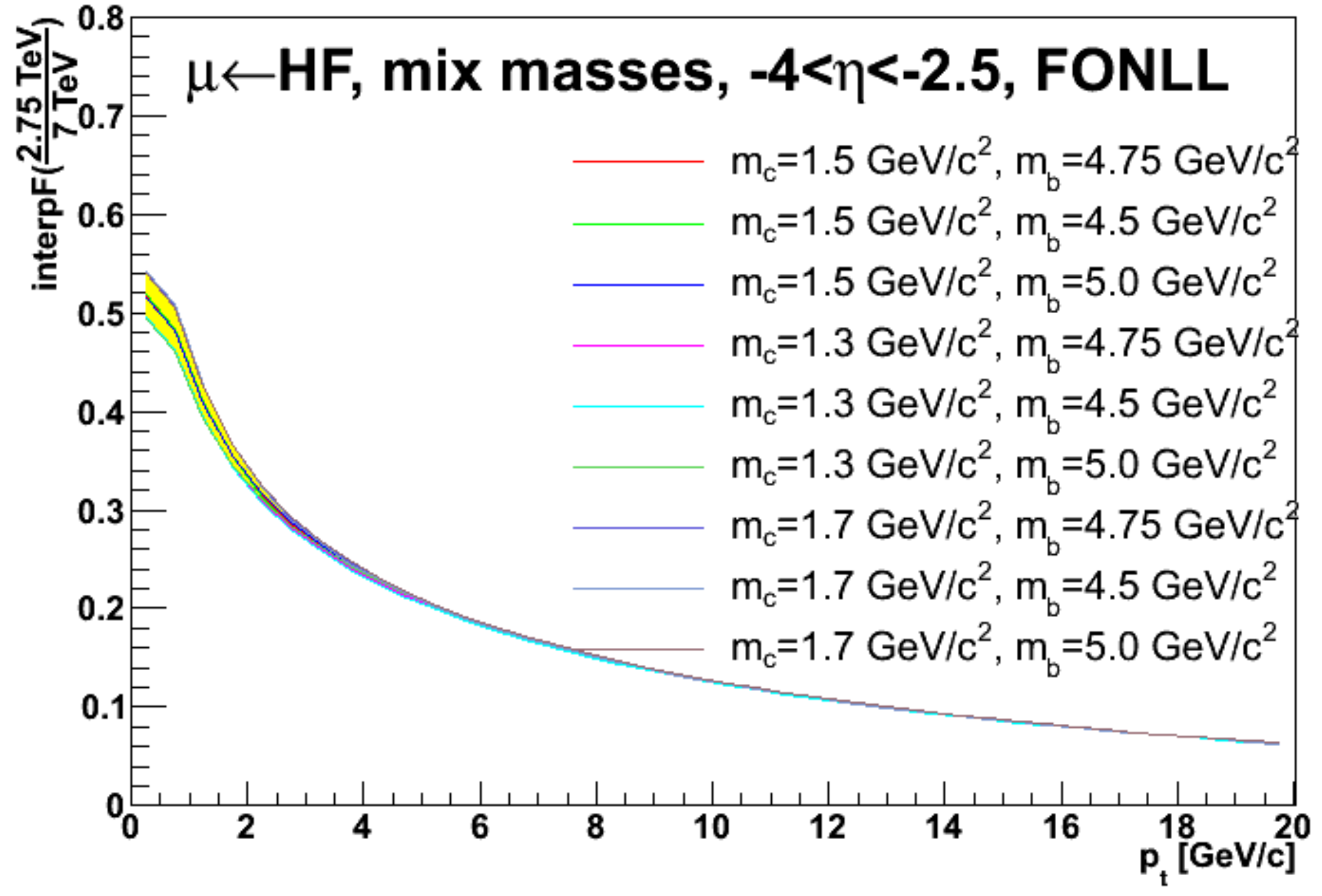}
        \includegraphics[width=0.479\columnwidth]
{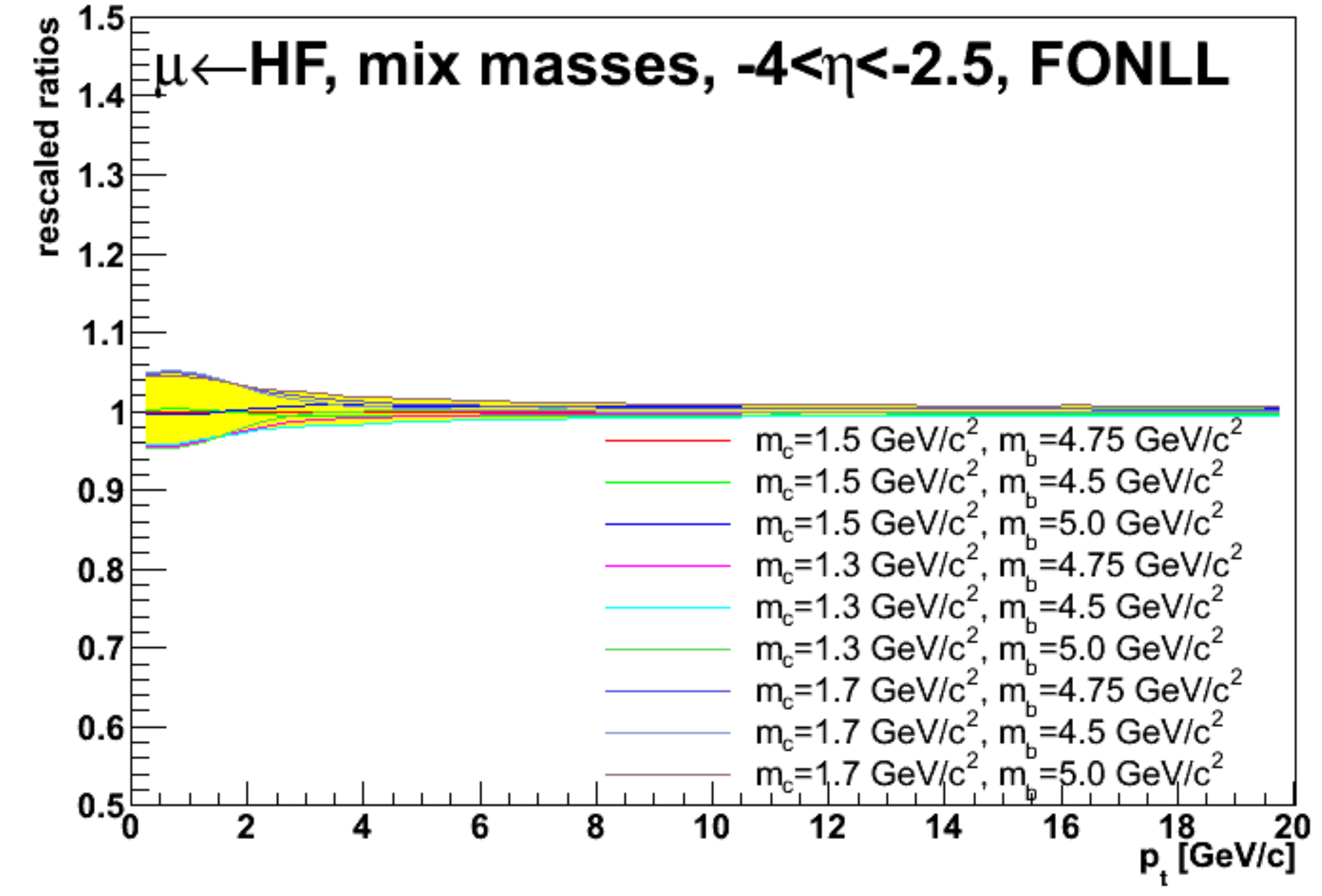}
\caption{Left: FONLL scaling factor from 7 TeV to 2.76 TeV for the 
measurement of $p_{\rm t}$-differential cross section of muons from heavy 
flavour decay with different combinations of quark masses indicated on the 
figure; right: corresponding 
relative systematic uncertainty.}
\label{fig:Muquarkmass}
\end{center}
\end{figure}

The FONLL scaling factor that will be applied to the $p_{\rm t}$-differential 
cross section of muons from heavy flavour decay 
measured in pp collisions at $\sqrt s$ = 7 TeV, in order to obtain the 
reference cross section at $\sqrt s$ = 2.76 TeV, was determined 
according to the procedure described in section~\ref{sec:recipe}. The only difference with 
respect to the case of electrons (section~\ref{sec:ElectronScaling}) is that two different rapidity 
ranges are used for the FONLL calculations: $|y|<0.8$ for electrons and 
$2.5<y<4$ for muons.  
Figure~\ref{fig:Muquarkmass} 
shows the scaling factor obtained by combining different 
sets of $c$ and $b$ quark masses and assuming that quark masses are 
unchanged at 2.76 TeV 
and 7 TeV. The scaling factor depends strongly on $p_{\rm t}$, in particular in 
the low $p_{\rm t}$ range ($p_{\rm t}<2$~GeV/$c$). It decreases from about 0.5 
to 0.2 in the $p_{\rm t}$ range 0--6~GeV/$c$ and tends to saturate 
(Fig.~\ref{fig:Muquarkmass}, left panel). The relative scaling factor, which 
gives the relative uncertainty, is depicted in the right panel of 
Fig.~\ref{fig:Muquarkmass}. Changes in the quark masses introduce a 
systematic uncertainty less than 
5$\%$ for $p_{\rm t}<2$~GeV/$c$, which can be neglected at higher $p_{\rm t}$ 
(range of interest for the measurement of the cross section of muons 
from heavy flavour decay). 
 
\begin{figure}[!htbp]
\begin{center}
        \includegraphics[width=0.479\columnwidth]
{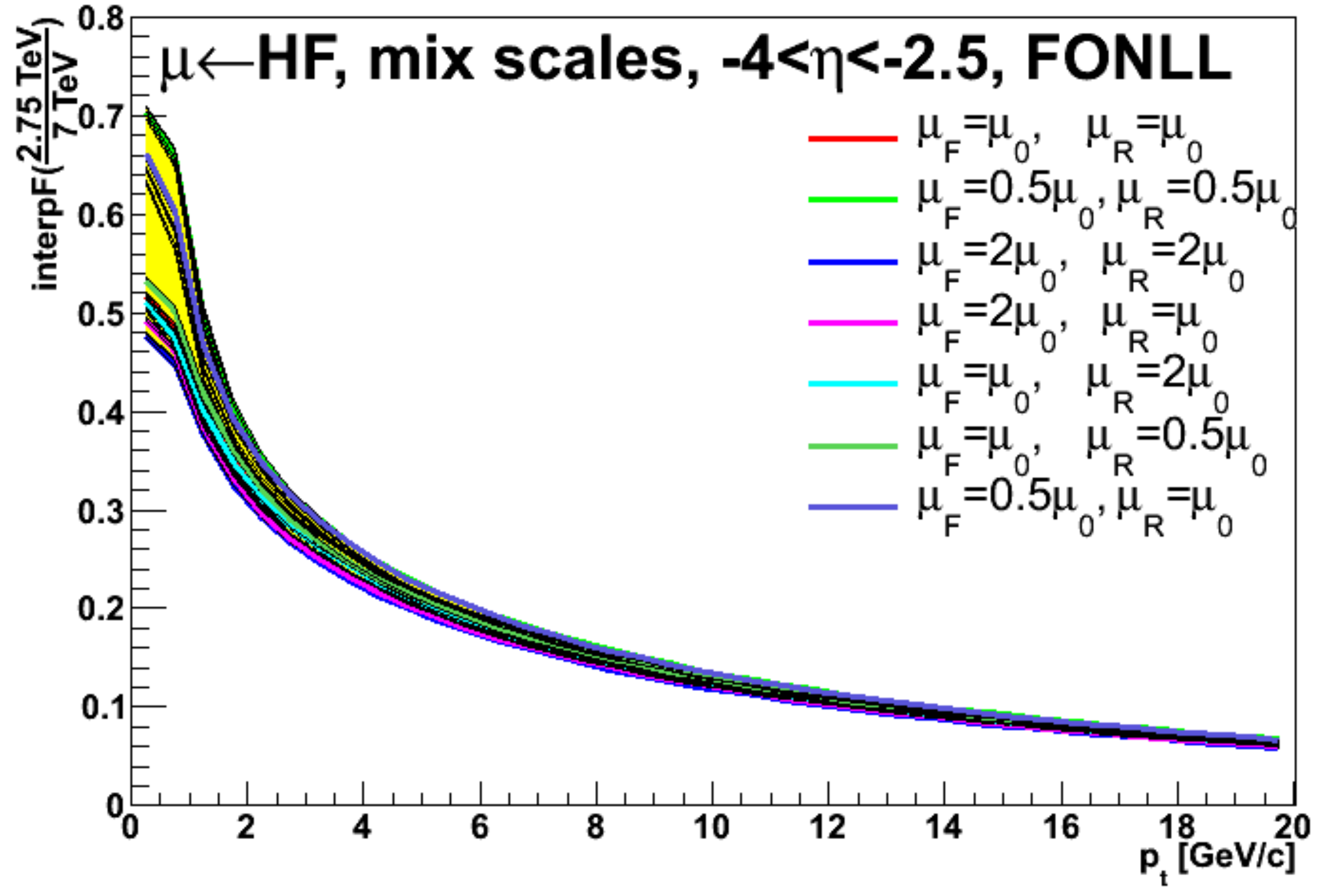}
        \includegraphics[width=0.479\columnwidth]
{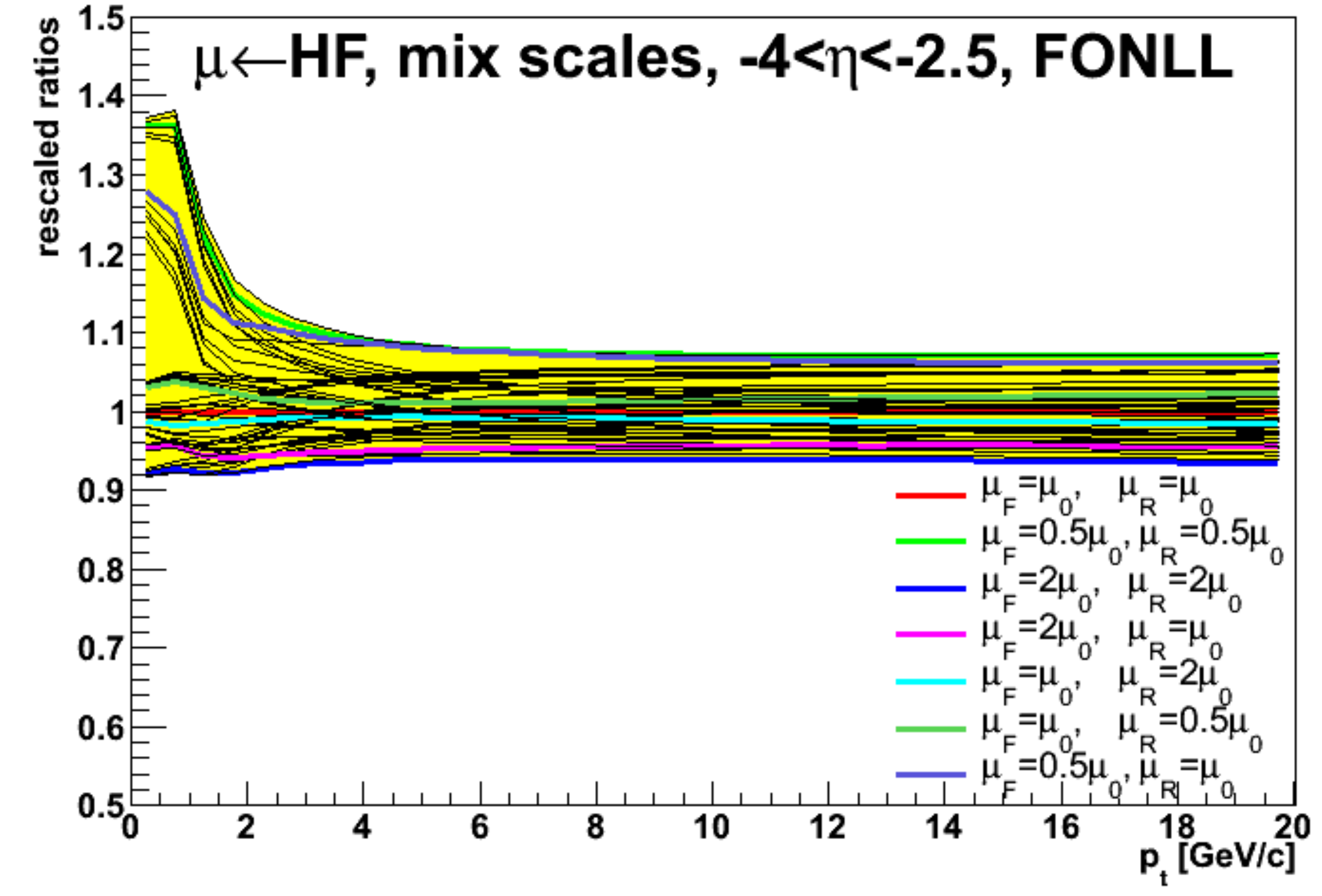}
\caption{Left: FONLL scaling factor from 7 TeV to 2.76 TeV for the 
measurement of $p_{\rm t}$ differential cross section of muons from heavy 
flavour decay with different combinations of QCD scales as indicated on the 
figure; right: corresponding 
relative systematic uncertainty. See the text for more detail.}
\label{fig:MuMixedScales}
\end{center}
\end{figure}

\begin{figure}[!htbp]
\begin{center}
         \includegraphics[width=0.479\columnwidth]
{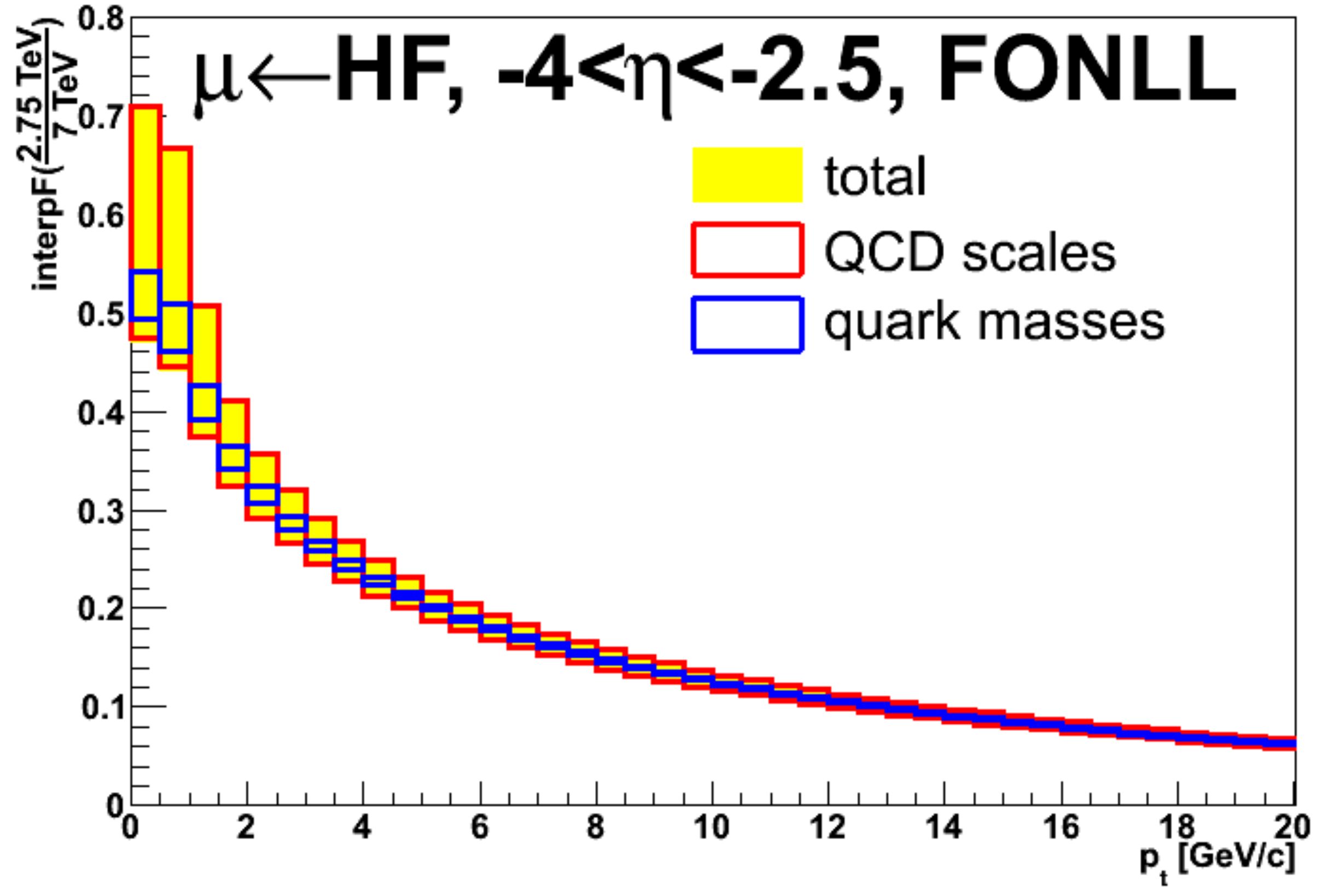}
        \includegraphics[width=0.479\columnwidth]     
{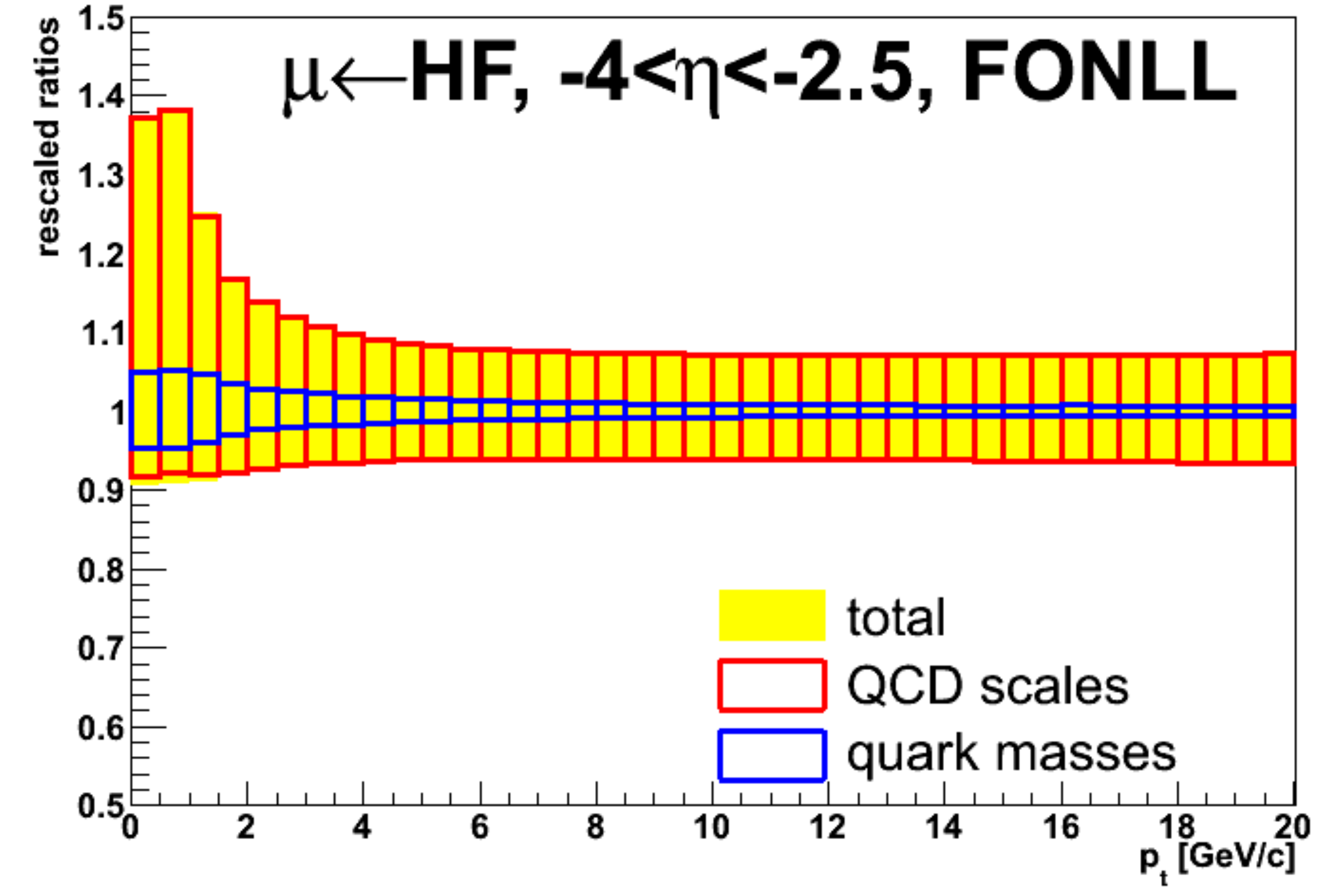}
\caption{Left: FONLL scaling factor from 7 TeV to 2.76 TeV for the 
measurement of $p_{\rm t}$ differential cross section of muons from heavy 
flavour decay with different combinations of QCD scales (red boxes) 
and quark masses (blue boxes). The yellow band is the total 
systematic uncertainty; right: corresponding relative systematic uncertainty.}
\label{fig:MuTotal}
\end{center}
\end{figure}

As in the other cases, it  was assumed that the pQCD scales do not change with $\sqrt{s}$. 
As for electrons at central rapidity, 
the influence of the pQCD scales variation on the FONLL scaling factor was 
investigated in two cases:
a) same scales for charm and beauty (correlated scales, 
colour lines in 
Fig.~\ref{fig:MuMixedScales});
b) different scales for charm and beauty (uncorrelated scales, black 
lines in Fig.~\ref{fig:MuMixedScales}).
Very similar results are 
obtained with correlated or uncorrelated scales 
for charm and beauty.  
At low muon $p_{\rm t}$ ($<2$~GeV/$c$) the uncertainty on the scaling factor reaches  
about $40\%$, while in the $p_{\rm t}> 2$~GeV/$c$ range it is 
below 10$\%$, independently of $p_{\rm t}$.

In summary, the FONLL scaling factor as a function of $p_{\rm t}$ 
obtained for different sets of quark masses (blue boxes) and 
pQCD scales (red boxes), as just discussed, is shown 
in Fig.~\ref{fig:MuTotal} (left panel).  The relative scaling factor is 
also shown in the right panel of the figure. For 
the systematic uncertainty from energy scaling, we consider the 
spread of the ratio obtained with the different sets of parameters 
(yellow band).

\begin{figure}[!htbp]
\begin{center}
  \includegraphics[width=0.575\columnwidth]{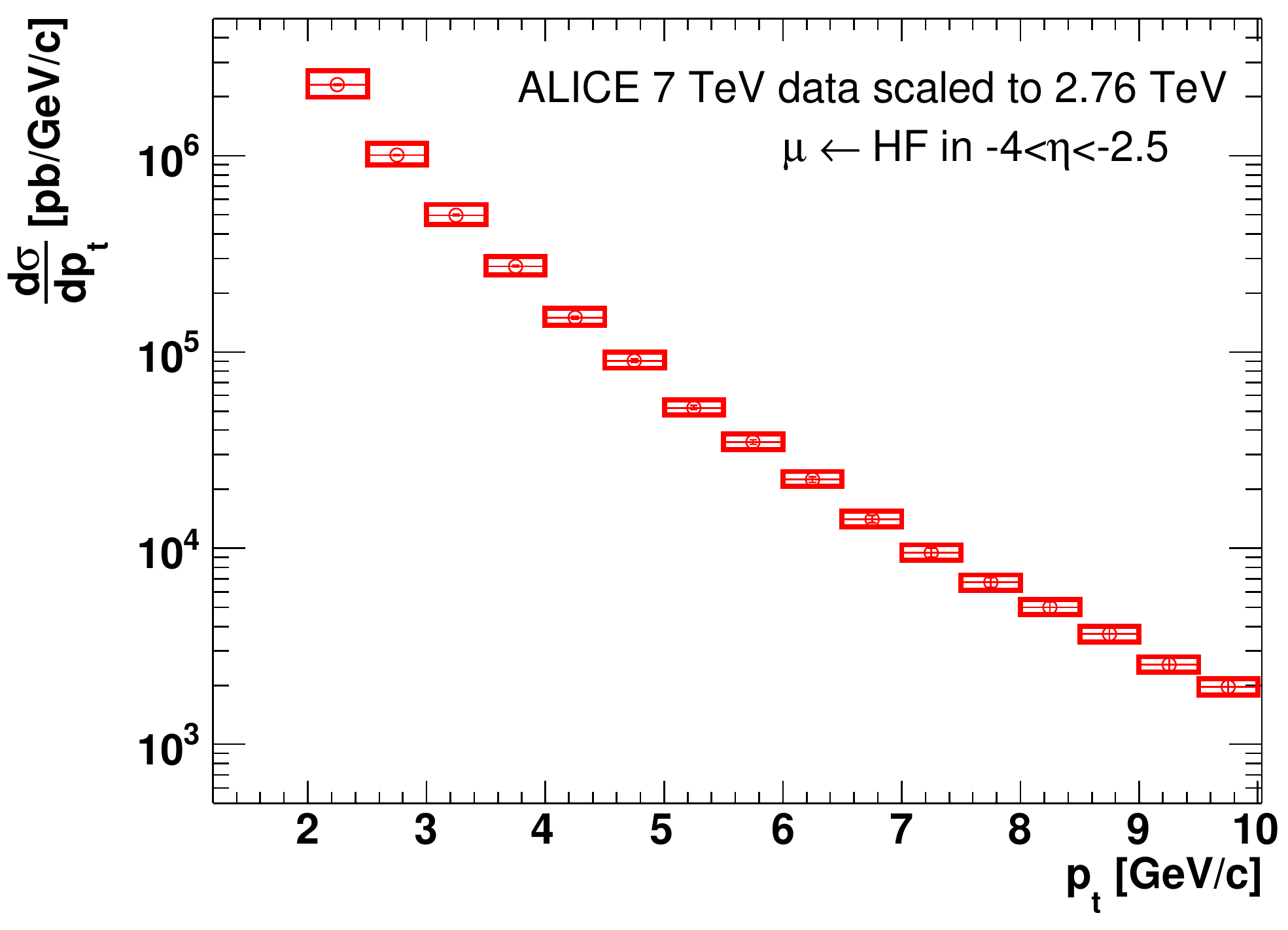}
\caption{$p_t$ differential cross section of muons from heavy flavour decays
obtained by scaling to $\sqrt{s}=2.76$~TeV 
the ALICE measurement at 7~TeV~\cite{XZnote}.}
\label{fig:muonsscaled}
\end{center}
\end{figure}

Figure~\ref{fig:muonsscaled} shows the cross section of muons from heavy flavour decays
obtained by scaling to $\sqrt{s}=2.76$~TeV 
the ALICE measurement at 7~TeV shown in Fig.~\ref{fig:SingleMuonPreliminary}.

\section{Conclusions} 
\label{sec:conclusions}

We have presented a procedure to define the $\sqrt s$-scaling factors for heavy flavour 
production cross sections in pp collisions at LHC energies. The scaling is based on 
perturbative QCD calculations, as implemented in the FONLL scheme, which
described reasonably well heavy flavour production as measured at the Tevatron and at the LHC.
For $D$ mesons and heavy flavour decay leptons, the scaling uncertainty from 7 to 2.76~TeV is of about 40\% for $p_{\rm t}<2$~GeV/$c$ and $<1$~GeV/$c$ respectively, and it decreases towards larger momenta, reaching a level below 10\%.
For $D$ mesons, the scaling was verified by comparing the scaled ALICE 7 TeV measurement 
to data by the CDF experiment at 1.96 TeV.

By applying the scaling to ALICE preliminary measurements at 7 TeV 
for $D^0$, $D^+$, $D^{*+}$, 
electrons and muons from heavy flavour decay, 
we have provided reference cross sections in pp collisions at 2.76 TeV.

\begin{acknowledgements}

The authors would like to thank M.~Cacciari for fruitful and stimulating discussions on the scaling procedure, as well as M.~Cacciari and H.~Spiesberger for providing the numerical cross sections from the FONLL and GM-VFNS calculations.

\end{acknowledgements}



\begin{thebibliography}{99}

\bibitem{PPRvol2}
 B. Alessandro {\it et al.} (ALICE Collaboration), J. Phys. {\bf G32} (2006) 1295.
 
 \bibitem{MNR}
 M. Mangano, P. Nason, and G. Ridolfi, Nucl. Phys. {\bf B373} (1992) 295.
 
 \bibitem{FONLL}
  M.~Cacciari, M.~Greco and P.~Nason,
  JHEP {\bf 9805} (1998) 007
  [arXiv:hep-ph/9803400];
  M.~Cacciari, S.~Frixione and P.~Nason,
  JHEP {\bf 0103} (2001) 006
  [arXiv:hep-ph/0102134].
  
\bibitem{CDFdata}
D. Acosta {\it et al.} (CDF Collaboration), Phys. Rev. Lett. {\bf 91} (2003) 241804.

\bibitem{Cacciari}
  M.~Cacciari, private communication.

\bibitem{GMVFNS}
  B.A. Kniehl {\it et al.}, Phys. Rev. Lett. {\bf 96} (2006) 012001; B.A. Kniehl, G.Kramer, I. Schienbein, and H. Spiesberger, private communication.


\bibitem{andrea}
A. Dainese {\it et al.} (ALICE Collaboration), proceedings of the XXII International Conference on Ultra-relativistic Nucleus--Nucleus Collisions, Quark Matter 2011, Annecy (2011).

\bibitem{renu}
R. Bala {\it et al.} (ALICE Collaboration), proceedings of the International Conference on the Physics and Astrophysics of the Quark Gluon Plasma, Goa (2010),
arXiv:1102.0199 [nucl-ex].

\bibitem{sigmaMB7}
K. Oyama {\it et al.} (ALICE Collaboration), proceedings of the XXII International Conference on Ultra-relativistic Nucleus--Nucleus Collisions, Quark Matter 2011, Annecy (2011).

\bibitem{noteHFE}
S. Masciocchi {\it et al.} (ALICE Collaboration), proceedings of the XXII International Conference on Ultra-relativistic Nucleus--Nucleus Collisions, Quark Matter 2011, Annecy (2011).

\bibitem{baines}
 J. Baines {\it et al.}, arXiv:hep-ph/0601164.
 
 \bibitem{vogt}
 M. Cacciari, P. Nason, and R. Vogt, Phys. Rev. Lett. {\bf 95} (2005) 122001.
 
\bibitem{XZnote}
X. Zhang {\it et al.} (ALICE Collaboration), proceedings of the XXII International Conference on Ultra-relativistic Nucleus--Nucleus Collisions, Quark Matter 2011, Annecy (2011).
  
\bibitem{pdg}
C. Amsler {\it et al.} (Particle Data Group), Phys. Lett. {\bf B667} (2008) 1.



  
\end{thebibliography}
\end{document}